\documentclass[12pt,fleqn,dvipdfm]{article}
\usepackage{graphicx}
\begin{document}
\baselineskip = 18 pt

\begin{center}
\Large \bf Computation in a single neuron: \\
Hodgkin and Huxley revisited
\end{center}

\bigskip\bigskip

\leftline{\large Blaise Ag\"uera~y~Arcas,$^1$ Adrienne L.
Fairhall,$^{2,3}$ and William Bialek$^{2,4}$}
\medskip
\leftline{$^1$Rare Books Library, Princeton University, Princeton,
New Jersey 08544 } \leftline{$^2$NEC Research Institute, 4
Independence Way, Princeton, New Jersey 08540}
\leftline{$^3$Department of Molecular Biology, Princeton
University, Princeton, New Jersey 08544} \leftline{$^4$Department
of Physics, Princeton University, Princeton, New Jersey 08544}

\leftline{\{blaisea,fairhall,wbialek\}@princeton.edu}
\bigskip

\leftline{31 December 2002}
\bigskip\bigskip

A spiking neuron ``computes'' by transforming a complex dynamical
input into a train of action potentials, or spikes. The
computation performed by the neuron can be formulated as
dimensional reduction, or feature detection, followed by a
nonlinear decision function over the low dimensional space.
Generalizations of the reverse correlation technique with white noise
input provide a numerical strategy for extracting the relevant low
dimensional features from experimental data, and information theory can be
used to evaluate the quality of the low--dimensional approximation. We
apply these methods to analyze the simplest biophysically
realistic model neuron, the Hodgkin--Huxley
model, using this system to illustrate the general methodological
issues. We focus on the features in the stimulus that trigger a
spike, explicitly eliminating the effects of interactions between
spikes. One can approximate this triggering ``feature space'' as a
two dimensional linear subspace in the high--dimensional space of input
histories, capturing in this way a substantial fraction of the mutual
information between inputs and spike time.   We find that an even better
approximation, however, is to describe the relevant subspace as
two dimensional, but curved; in this way we can capture 90\% of the
mutual information even at high time resolution.   Our analysis provides
a new understanding of the computational properties of the Hodgkin--Huxley
model. While it is common to approximate neural behavior as ``integrate
and fire,'' the HH model is not an integrator nor is it well described by
a single threshold.


\vfill\newpage
\section{Introduction}

\label{s:introduction}

On short timescales, one can conceive of a single neuron as a
computational device that maps inputs at its synapses into a
sequence of action potentials or spikes.  To a good approximation,
the dynamics of this mapping are determined by the kinetic
properties of ion channels in the neuron's membrane.  In the fifty
years since the pioneering work of Hodgkin and Huxley,  we have
seen the evolution of an ever more detailed description of channel
kinetics, making it plausible that the short time dynamics of
almost any neuron we encounter will be understandable in terms of
interactions among a   mixture of diverse but known channel types
\cite{hille,koch}. The existence of so nearly complete a
microscopic picture of single neuron dynamics brings into focus a
very different question: What does the neuron compute?  Although
models in the Hodgkin--Huxley tradition define a dynamical system
which will reproduce the behavior of the neuron, this description
in terms of differential equations is far from our intuition
about---or the formal description of---computation.

The problem of what neurons compute is one instance of a more
general problem in modern quantitative biology and biophysics:
Given a progressively more complete microscopic description of
proteins and their interactions, how do we understand the
emergence of function?  In the case of neurons, the proteins  are the ion
channels, and the interactions are very simple---current flows through
open channels, charging the cell's capacitance, and all channels
experience the resulting voltage. Arguably there is no other network of
interacting proteins for which the relevant equations are known in such
detail; indeed some efforts to understand function and computation in
other networks of proteins make use of analogies to neural systems
\cite{bray}. Despite the relative completeness of our microscopic picture
for neurons, there remains a huge gap between the description of
molecular kinetics and the understanding of function.  Given some
complex dynamic input to a neuron, we might be able to simulate
the spike train that will result, but we are hard pressed to look
at the equations for channel kinetics and say that this
transformation from inputs to spikes is equivalent to some simple
(or perhaps not so simple) computation such as filtering,
thresholding, coincidence detection or feature extraction.

Perhaps the problem of understanding computational function in a model
of ion channel dynamics is a symptom of a much deeper mathematical
difficulty.  Despite the fact that all computers are dynamical systems,
the natural mathematical objects in dynamical systems theory are very
different from those in the theory of computation, and it is not clear
how to connect these different formal schemes. Finding a general mapping
from dynamical systems to their equivalent computational functions is a
grand challenge, but we will take a more modest approach.

We believe that a key intuition for understanding neural
computation is the concept of feature selectivity:  While the
space of inputs to a neuron---whether we think of inputs as
arriving at the synapses or being driven by sensory signals
outside the brain---is vast, individual neurons are sensitive only
to some restricted set of features in this vast space. The most
general way to formalize this intuition is to say that we can
compress (in the information theoretic sense) our description of
the inputs without losing any information about the neural output
\cite{bottleneck}.  We might hope that this selective compression
of the input data has a simple geometric description, so that the
relevant bits about the input correspond to coordinates along some
restricted set of relevant dimensions in the space of inputs.  If
this is the case, feature selectivity should be formalized as a
reduction of dimensionality \cite{bialek88}, and this is the
approach we follow here.  Closely related work on the use of
dimensionality reduction to analyze neural feature selectivity has
been described in recent papers \cite{bill&robinprep,tatyana}.

Here we develop the idea of dimensionality reduction as a tool for
analysis of neural computation, and apply these tools to the
Hodgkin--Huxley model.  While our initial goal was to test new
analysis methods in the context of a presumably simple and well
understood model, we have found that the Hodgkin--Huxley neuron
performs a computation of surprising richness.  Preliminary
accounts of these results have appeared \cite{baathesis,nips2000}.

\section{Dimensionality reduction}

Neurons take input signals at their synapses and give as output a
sequences of  spikes.  To characterize a neuron completely is to
identify the mapping between neuronal input and the spike train
the neuron produces in response. In the absence of any simplifying
assumptions, this  requires probing the system with every possible
input. Most often, these inputs are spikes from other neurons;
each neuron typically has of order $N\sim 10^3$ presynaptic
connections.  If the system operates at 1 msec resolution and the
time window of relevant inputs is 40 msec, then we can think of a
single neuron as having an input described by a $\sim 4\times
10^4$ bit word---the presence or absence of a spike in each 1 msec
bin for each presynaptic cell---which is then mapped to a one
(spike) or zero (no spike). More realistically, if average spike
rates are $\sim 10 \,{\rm s}^{-1}$, the input words can be
compressed by a factor of ten.   In this picture, a neuron
computes a Boolean function over roughly 4000 variables. Clearly
one cannot sample every one of the $\sim 2^{4000}$ inputs to
identify the neural computation. Progress requires making some
simplifying assumption about the function computed by the neuron
so that we can vastly reduce the space of possibilities over which
to search.  We use the idea of dimensionality reduction in this
spirit, as a simplifying assumption which allows us to make
progress but which also must be tested directly.

The ideas of feature selectivity and dimensionality reduction have
a long history in neurobiology. The idea of receptive fields as
formulated by Hartline, Kuffler, and Barlow for the visual system
gave a picture of neurons as having a template against which
images would be correlated \cite{hartline,kuffler,barlow53}. If we
think of images as vectors in a high dimensional space, with
coordinates determined by the intensities of each pixel, then the
simplest receptive field models describe the neuron as sensitive
to only one direction or projection in this high dimensional
space.  This picture of projection followed by thresholding or
some other nonlinearity to determine the probability of spike
generation was formalized in the linear perceptron
\cite{rosen2,rosenblatt}. In subsequent work Barlow and others
\cite{barlow2} characterized neurons in which the receptive field
has subregions in space and time such that summation is at least
approximately linear in each subregion but these summed signals
interact nonlinearly, for example to generate direction
selectivity and motion sensitivity. We can think of Hubel and
Wiesel's description of complex and hypercomplex cells
\cite{hubel&wiesel62} again as a picture of approximately linear
summation within subregions followed by nonlinear operations on
these multiple summed signals. More formally,  the proper
combination of linear summation and nonlinear or logical
operations may provide a useful bridge from receptive field
properties to proper geometric primitives in visual computation
\cite{iverson&zucker}. In the same way that a single receptive
field or perceptron model has one relevant dimension in the space
of visual stimuli, these more complex cells have as many relevant
dimensions as there are independent subregions of the receptive
field.  While this number is larger than one, it still is much
smaller than the full dimensionality of the possible
spatiotemporal variations in visual inputs.

The idea that neurons in the auditory system might be described by
a filter followed by a nonlinear transformation to determine the
probability of spike generation was the inspiration for de Boer's
development  \cite{deBoer} of triggered or reverse correlation.
Modern uses of reverse correlation to characterize the filtering
or receptive field properties of a neuron often emphasize that
this approach provides a ``linear approximation'' to the
input/output properties of the cell, but the original idea was
almost the opposite:  Neurons clearly are nonlinear devices, but
this is separate from the question of whether the probability of
generating a spike is determined by a simple projection of the
sensory input onto a single filter or template.  In fact, as
explained by Rieke et al. (1997), linearity is seldom a good
approximation for the neural input/output relation, but if there
is one relevant dimension then (provided that input signals are
chosen with suitable statistics) the reverse correlation method is
guaranteed to find this one special direction in the space of
inputs to which the neuron is sensitive.  While the reverse
correlation method is guaranteed to find the one relevant
dimension if it exists, the method does not include any way of
testing for other relevant dimensions, or more generally for
measuring the dimensionality of the relevant subspace.

The idea of characterizing neural responses directly as the
reduction of dimensionality emerged from studies   \cite{bialek88} of a
motion sensitive neuron in the fly visual system. In
particular, this work suggested that it is possible to estimate
the dimensionality of the relevant subspace, rather than just
assuming that it is small (or equal to one).  More recent work on
the fly visual system has exploited the idea of dimensionality
reduction  to probe both the structure and adaptation of the
neural code \cite{naama,anature} and the nature of the computation
which extracts the motion signal from the spatiotemporal array of
photoreceptor inputs \cite{bill&robinprep}.  Here we review the
ideas of dimensionality reduction from previous work; extensions
of these ideas begin in Section 3.

In the spirit of neural network models we will
simplify away the spatial structure of neurons and consider time dependent
currents $I(t)$  injected into a point--like neuron.
While this misses much of the complexity of real cells, we will find
that even this system is highly nontrivial. If the input is an
injected current, then the neuron maps the history of this
current, $I(t < t_0)$, into the presence or absence of a spike at
time $t_0$. More generally, we might imagine that the cell (or our
description) is noisy, so that there is a probability of spiking
$P[{\rm spike \, at \,} t_0 | I(t < t_0)]$ which depends on the
current history. The dependence on the {\em history} of the
current means that the input signal still is  high dimensional,
even without spatial dependence. Working at time resolution
$\Delta t$ and assuming that currents in a window of size $T$ are
relevant to the decision to spike, the input space is of dimension
$D = T/\Delta t$, where $D$ is often of order 100.

The idea of dimensionality reduction is that the probability of
spike generation is sensitive only to some limited number of
dimensions $K$ within the $D$ dimensional space of inputs.  We
begin our analysis by searching for linear subspaces, that is a set of
signals $s_1 , s_2, \cdots , s_K$ which can be constructed by
filtering the current,
\begin{equation}
s_\mu = \int_0^\infty dt f_\mu (t) I(t_0 - t) ,
\end{equation}
so that the probability of spiking depends only on this small set
of signals,
\begin{equation}
P[{\rm spike \, at \,} t_0 | I(t < t_0)] = P[{\rm spike \, at \,}
t_0]g(s_1, s_2, \cdots , s_K ), \label{e:Kprojs}
\end{equation}
where the inclusion of the average probability of spiking, $P[{\rm
spike \, at \,} t_0]$, leaves $g$ dimensionless. If we think of
the current $I(t_0-T < t < t_0)$ as a $D$ dimensional vector, with
one dimension for each discrete sample at spacing $\Delta t$, then
the filtered signals $s_i$ are linear projections of this vector.
In this formulation, characterizing the computation done by a
neuron involves three steps:
\begin{enumerate}
\item Estimate the number of relevant
stimulus dimensions $K$, hopefully much less than the original
dimensionality $D$.
\item Identify a set of filters which project into
this relevant subspace.
\item Characterize the nonlinear
function $g(\vec s )$.
\end{enumerate}
The classical perceptron--like cell of neural network theory would
have only one relevant dimension, given by the vector of weights,
and a simple form for $g$, typically a sigmoid.

Rather than trying to look directly at the distribution of
spikes given stimuli, we follow de Ruyter van Steveninck
and Bialek (1988) and consider the  distribution of signals conditional on
the response,
$P[I(t < t_0)| {\rm spike \, at \, } t_0  ]$, also called the response
conditional ensemble (RCE); these are related by Bayes' rule,
\begin{equation}
{{P[{\rm spike \, at \,} t_0 | I(t < t_0)]}\over{P[{\rm spike \,
at \, } t_0 ]}} = {{P[I(t < t_0)| {\rm spike \, at \,} t_0
]}\over{P[I(t < t_0)]}} \,. \label{e:Bayes}
\end{equation}
We can now compute various moments of the RCE.  The first
moment is the spike triggered average stimulus (STA),
\begin{equation} \label{e:sta} {\rm STA}(\tau ) = \int [dI]
\,P[I(t < t_0)| {\rm spike \, at \,} t_0  ] I(t_0 -\tau) \, ,
\end{equation}
which  is the object that one computes in reverse correlation
\cite{deBoer,spikes}.   If we choose the distribution of input
stimuli $P[I(t < t_0)]$ to be Gaussian white noise, then for a
perceptron--like neuron sensitive to only one direction in
stimulus space, it can be shown that the STA or first moment of
the response conditional ensemble is proportional to the vector or
filter $f(\tau )$ that defines this direction \cite{spikes}.

Although it is a theorem that the STA is proportional to the
relevant filter $f(\tau)$, in principle it is possible that the
proportionality constant is zero, most plausibly if the neuron's
response has some symmetry, such as phase invariance in the
response of high frequency auditory neurons.  It also is worth
noting that what is really important in this analysis is the
Gaussian distribution of the stimuli, not the ``whiteness'' of the
spectrum.  For nonwhite but Gaussian inputs the STA measures the
relevant filter blurred by the correlation function of the inputs
and hence the true filter can be recovered (at least in principle)
by deconvolution.   For nonGaussian signals and nonlinear neurons
there is no corresponding guarantee that the selectivity of the
neuron can be separated from correlations in the stimulus
\cite{tatyana}.

To obtain more than one relevant direction (or to reveal relevant
directions when symmetries cause the STA to vanish), we proceed to
second order and compute the covariance matrix of fluctuations
around the spike triggered average,
\begin{equation}
\label{e:cov} C_{\rm spike} (\tau , \tau ') = \int [dI] \,P[I(t <
t_0)| {\rm spike \, at \,} t_0  ] I(t_0 -\tau) I(t_0 -\tau') -
{\rm STA}(\tau ) {\rm STA}(\tau ') .
\end{equation}
In the same way that we compare the spike triggered average to
some constant average level of the signal in the whole experiment,
we  compare the covariance matrix $C_{\rm spike}$ with the
covariance of the signal averaged over the whole experiment,
\begin{equation}
C_{\rm prior} (\tau , \tau ')= \int [dI] \,P[I(t < t_0)]  I(t_0
-\tau) I(t_0 -\tau') ,
\end{equation}
to construct the change in the covariance matrix
\begin{equation}
\label{e:deltac} \Delta C = C_{\rm spike} - C_{\rm prior} \, .
\end{equation}
With time resolution $\Delta t$ in a window of duration $T$ as above, all
of these covariances are
$D\times D$ matrices.   In the same way that the spike triggered average
has the clearest interpretation when we choose inputs from a Gaussian
distribution, $\Delta C$ also has the clearest interpretation in
this case.  Specifically, if inputs are drawn from a Gaussian
distribution then it can be shown that \cite{bill&robinprep}:
\begin{enumerate}
\item If the neuron is sensitive to a limited set of $K$ input
dimensions as in Eq. (\ref{e:Kprojs}), then $\Delta C$ will have
only $K$ nonzero eigenvalues.\footnote{As with the STA, it is in
principle possible that symmetries or accidental features of the
function $g({\vec s})$ would cause some of the $K$ eigenvalues to
vanish, but this is very unlikely.}  In this way we can measure
directly the dimensionality $K$ of the relevant subspace.
\item If the distribution of inputs are both Gaussian and white, then the
eigenvectors associated with the nonzero eigenvalues span the same space
as that spanned by the filters $\{ f_\mu (\tau ) \}$.
\item  For nonwhite (correlated) but still Gaussian inputs, the
eigenvectors span the space of the filters $\{ f_\mu (\tau ) \}$
blurred by convolution with the correlation function of the inputs.
\end{enumerate}
Thus the analysis of $\Delta C$ for neurons responding to Gaussian
inputs should allow us to identify the subspace of inputs of
relevance and to test specifically the hypothesis that this
subspace is of low dimension.

Several points are worth noting.  First, except in special cases,
the eigenvectors of $\Delta C$ and the filters $\{ f_\mu (\tau )
\}$ are {\em not} the principal components of the response
conditional ensemble, and hence this analysis of $\Delta C$ is not
a principal component analysis.  Second, the nonzero eigenvalues
of $\Delta C$ can be either positive or negative, depending on
whether the variance of inputs along that particular direction is
larger or smaller in the neighborhood of a spike. Third, although
the eigenvectors span the relevant subspace, these eigenvectors do
not form a preferred coordinate system within this subspace.
Finally, we emphasize that dimensionality
reduction---identification of the relevant subspace---is only the
first step in our analysis of the computation done by a neuron.

\section{Measuring the success of dimensionality reduction}

\label{s:info}

The claim that certain stimulus features
are most relevant is in effect a model for the neuron, so the next
question is how to measure the effectiveness or accuracy of this
model. Several different ideas have been suggested in the
literature as ways of testing  models based on linear
receptive fields in the visual system \cite{dan,markusmodel} or
linear spectrotemporal receptive fields in the auditory system
\cite{theunissen}. These methods have in common that they
introduce a  metric to measure performance---for example,
mean square error in predicting the firing rate as  averaged over
some window of time. Ideally we would like to have a performance
measure which avoids any arbitrariness in the choice of metric,
and such metric free measures are provided uniquely by information
theory \cite{shannon,cover&thomas}.

Observing the arrival time $t_0$ of a single spike provides a
certain amount of information about the  input signals. Since
information is mutual, we can also say that knowing the input
signal trajectory $I(t < t_0)$ provides information about the
arrival time of the  spike.  If ``details are irrelevant'' then we
should be able to discard these details from our description of
the stimulus and yet preserve the mutual information between the
stimulus and spike arrival times; for an abstract discussion of
such selective compression see \cite{bottleneck}.  In constructing
our low dimensional model, we represent the complete ($D$
dimensional) stimulus $I(t < t_0)$ by a smaller number ($K < D$)
of dimensions ${\vec s} = (s_1, s_2, \cdots, s_K)$.

The mutual information ${\cal I} [I(t < t_0) ; t_0]$ is a property
of the neuron itself, while the mutual information ${\cal I}
[{\vec s} ; t_0]$ characterizes how much our reduced description
of the stimulus can tell us about when spikes will occur.
Necessarily our reduction of dimensionality causes a loss of
information, so that
\begin{equation}
{\cal I} [{\vec s}; t_0] \leq {\cal I} [I(t < t_0) ; t_0] ,
\end{equation}
but if our reduced description really captures the computation
done by the neuron then the two information measures will be very
close.  In particular if the neuron were described exactly by a
lower dimensional model---as for a linear perceptron, or for an
integrate and fire neuron \cite{i&f}---then the two information
measures would be equal.  More generally, the ratio ${\cal I}
[{\vec s}; t_0] /{\cal I} [I(t < t_0) ; t_0]$ quantifies the
efficiency of the low dimensional model, measuring the fraction of
information about spike arrival times that our $K$ dimensions
capture from the full signal $I(t<t_0)$.

As shown by Brenner et al. (2000), the arrival time of a
single spike provides an information
\begin{equation}
{\cal I} [I(t < t_0) ; t_0] \equiv I_{\rm one\, spike} = {1\over
T}\int_0^T dt\, {{r(t)}\over{\bar r}}
\log_2\left[ {{r(t)}\over{\bar r}}\right] ,
\label{e:onespikeinfo}
\end{equation}
where $r(t)$ is the time dependent spike rate, $\bar r$ is the
average spike rate, and $\langle \cdots\rangle$ denotes an average
over time. In principle, information should be calculated as an
average over the distribution of stimuli, but the  ergodicity of
the stimulus justifies replacing this ensemble average with a time
average. For a deterministic system like the Hodgkin--Huxley
equations, the spike rate is a singular function of time:  given
the inputs $I(t)$, spikes occur at definite times with no
randomness or irreproducibility. If we observe these responses
with a time resolution $\Delta t$, then for $\Delta t$
sufficiently small the rate $r(t)$ at any time $t$ either is zero
or corresponds to a single spike occurring in one bin of size
$\Delta t$, that is $r = 1/\Delta t$.  Thus the information
carried by a single spike is
\begin{equation}
I_{\rm one \, spike} = -\log_2 {\bar r}\Delta t \, .
\label{e:onespikemax}
\end{equation}
On the other hand, if the probability of spiking really depends
only on the stimulus dimensions $s_1, s_2, \cdots, s_K$, we can
substitute
\begin{equation}
{{r(t)}\over{\bar r}} \rightarrow {{P({\vec s} | {\rm spike \, at
\,}t)}\over{P({\vec s})}} \, . \label{e:bayes-spikes}
\end{equation}
Replacing the time averages in Eq.~(\ref{e:onespikeinfo}) with
ensemble averages, we find
\begin{equation}
{\cal I} [{\vec s}; t_0] \equiv I_{\rm one \, spike}^{\vec s} =
\int d^K s \, P({\vec s} | {\rm spike \,
at \,}t) \log_2\left[ {{P({\vec s} | {\rm spike \, at
\,}t)}\over{P({\vec s} )}}\right] \, ; \label{e:infoin2d}
\end{equation}
for details of these arguments see \cite{naama-info-method}.
This allows us to compare the information captured by the
$K$ dimensional reduced model with the true information carried by single
spikes in the spike train.

For reasons that we will discuss in the following section, and as
was pointed out in \cite{nips2000,i&f}, we will be considering
{\em isolated} spikes, i.e. spikes separated from previous spikes
by a period of silence. This has important consequences for our
analysis. Most significantly, as we will be considering spikes
that occur on a background of silence, the relevant stimulus
ensemble, conditioned on the silence, is no longer Gaussian.
Further, we will need to refine our information estimate.

The derivation of Eq. (\ref{e:onespikeinfo}) makes clear that a
similar formula must determine the information carried by the
occurrence time of any event, not just single spikes:  we can
define an event rate in place of the spike rate, and then
calculate the information carried by these events
\cite{naama-info-method}.  In the present case we wish to compute
the information obtained by observing an isolated spike, or
equivalently by the event silence+spike. This is
straightforward---we replace the spike rate by the rate of
isolated spikes and Eq. (\ref{e:onespikeinfo})  will give us the
information carried by the arrival time of a single isolated
spike.  The problem is that this information includes both the
information carried by the occurrence of the spike and the
information conveyed in the condition that there were no spikes in
the preceding $t_{\rm silence}$ msec [for an early discussion of
the information carried by silence see de Ruyter van Steveninck
and Bialek (1988)].  We would like to separate these
contributions, since our idea of dimensionality reduction applies
only the triggering of a spike, not to the temporally extended
condition of nonspiking.

 To separate the information carried by the isolated
spike itself, we have to ask how much information we gain by
seeing an isolated spike given that the condition for isolation
has already been met. As discussed by Brenner et al. (2000b), we
can compute this information by thinking about the distribution of
times at which the isolated spike can occur.  Given that we know
the input stimulus, the distribution of times at which a single
isolated spike will be observed is proportional to $r_{\rm
iso}(t)$, the time dependent rate or peri--stimulus time histogram
for isolated spikes.  With proper normalization we have
\begin{equation}
P_{\rm iso} (t|{\rm inputs}) = {1\over T} \cdot {1\over {\bar r}_{\rm
iso}} r_{\rm iso}(t) ,
\end{equation}
where $T$ is duration of the (long) window in which we can look for the
spike and ${\bar r}_{\rm iso}$ is the average rate of isolated spikes.
This distribution has an entropy
\begin{eqnarray}
S_{\rm iso} (t|{\rm inputs}) &=& -\int_0^T dt\, P_{\rm iso} (t|{\rm
inputs})  \log_2 P_{\rm iso} (t|{\rm inputs}) \\
&=& -{1\over T} \int_0^T dt\, {{r_{\rm iso}(t)}\over{{\bar r}_{\rm iso}}}
\log_2\left[ {1\over T} \cdot {{r_{\rm iso}(t)}\over{{\bar r}_{\rm
iso}}}\right]\\
&=&
\log_2(T {\bar r}_{\rm iso} \Delta t ) \,{\rm bits,}
\end{eqnarray}
where again we use the fact that for a deterministic system the time
dependent rate must be either zero or the maximum allowed by our time
resolution $\Delta t$.  To compute the information carried by a single
spike we need to compare this entropy with the total entropy possible
when we don't know the inputs.

It is tempting to think that without knowledge of the inputs, an
isolated spike is equally likely to occur anywhere in the window
of size $T$, which leads us back to  Eq. (\ref{e:onespikemax}) with $\bar
r$ replaced by ${\bar r}_{\rm iso}$.  In this case, however, we are
assuming that the condition for isolation has already been met.  Thus,
even without observing the inputs, we know that isolated spikes can occur
only in windows of time whose total length is $T_{\rm silence} =
T\cdot P_{\rm silence}$, where
$P_{\rm silence}$ is the probability that any moment in time is at
least $t_{\rm silence}$ after the most recent spike.  Thus the
total entropy of isolated spike arrival times (given that the
condition for silence has been met) is reduced from $\log_2 T$ to
\begin{equation}
S_{\rm iso} (t|{\rm silence}) = \log_2( T\cdot P_{\rm silence}),
\end{equation}
and the information that the spike carries beyond what we know from the
silence itself is
\begin{eqnarray}
\Delta I_{\rm iso\, spike} &=& S_{\rm iso} (t|{\rm silence}) - S_{\rm iso}
(t|{\rm inputs})  \\
&=& {1\over T} \int_0^T dt\, {{r_{\rm iso}(t)}\over{{\bar r}_{\rm iso}}}
\log_2\left[   {{r_{\rm iso}(t)}\over{{\bar r}_{\rm
iso}}} \cdot P_{\rm silence} \right]
\label{e:iso-spikeinfo2} \\
&=&  -\log_2\left({\bar r_{\rm iso}\Delta
t}\right) + \log_2 P_{\rm silence} \,\, {\rm bits.}
\label{e:iso-spikeinfo3}
\end{eqnarray}
This information, which is defined independently of any model for
the feature selectivity of the neuron, provides the benchmark
against which our reduction of dimensionality  will be measured.
To make the comparison, however, we need the analog of Eq.
(\ref{e:infoin2d}).

Equation (\ref{e:iso-spikeinfo2}) provides us with an expression
for the information conveyed by isolated spikes in terms of the
probability that these spikes occur at particular times; this is
analogous to Eq. (\ref{e:onespikeinfo}) for single (nonisolated)
spikes.  If we follow a path analogous to that which leads from   Eq.
(\ref{e:onespikeinfo})  to Eq. (\ref{e:infoin2d}) we find
an expression for the information which an isolated spike provides
about the $K$ stimulus dimensions ${\vec s}$:
\begin{eqnarray}
\Delta I_{\rm iso\, spike}^{\vec s} &=& \int d{\vec s} \, P({\vec
s} | {\rm iso~spike~at~} t) \log_2\left[ {{P({\vec s} | {\rm
iso~spike~at~} t) } \over{P({\vec s} | {\rm silence}) }} \right]
\nonumber\\
&& +\langle \log_2 P({\rm silence} |{\vec s}) \rangle ,
\label{e:2diso1}
\end{eqnarray}
where the prior is now also conditioned on silence: $P({\vec
s}|{\rm silence})$, is the distribution of ${\vec s}$ given that
${\vec s}$ is preceded by a silence of at least $t_{\rm silence}$.
Notice that this silence conditioned distribution is not knowable
{\em a priori}, and in particular it is not Gaussian; $P({\vec
s}|{\rm silence})$ must be sampled from data.

The last term in Eq. (\ref{e:2diso1}) is the entropy of a binary
variable which indicates whether particular moments in time are
silent given knowledge of the stimulus.  Again, since the HH model
is deterministic, this conditional entropy should be zero if we
keep a complete description of the stimulus.  In fact we're not
interested in describing those features of the stimulus which lead
to silence, nor is it fair (as we will see) to judge the success
of dimensionality reduction by looking at the prediction of
silence which necessarily involves multiple dimensions.   To make
a meaningful comparison, then, we will assume that there is a
perfect description of the stimulus conditions leading to silence,
and focus on the stimulus features that trigger the isolated
spike. When we approximate these features by the $K$ dimensional
space ${\vec s}$, we capture an amount of information
\begin{equation}
\Delta I_{\rm iso\, spike}^{\vec s}  = \int d{\vec s} \, P({\vec
s} | {\rm iso~spike~at~} t) \log_2\left[ {{P({\vec s} | {\rm
iso~spike~at~} t) } \over{P({\vec s} | {\rm silence}) }} \right] .
\label{e:2diso2}
\end{equation}
This is the information which we can compare with $\Delta I_{\rm iso\,
spike}$ in Eq. (\ref{e:iso-spikeinfo3}) to determine the efficiency of our
dimensionality reduction.

\section{Characterising the Hodgkin-Huxley neuron}

For completeness we begin with a brief review of the dynamics of
the space--clamped Hodgkin--Huxley neuron \cite{hh52}. Hodgkin and
Huxley modelled the dynamics of the current through a patch of
membrane with ion--specific conductances:
\begin{equation} \label{e:model}
C{{dV}\over{dt}} = I(t) - \bar{g}_{\rm K}n^4\left(V-V_{\rm
K}\right) - \bar{g}_{\rm Na}m^3h\left(V-V_{\rm Na}\right) -
\bar{g}_l\left(V-V_l\right),
\end{equation}
where $I(t)$ is injected current, ${\rm K}$ and ${\rm Na}$
subscripts denote potassium-- and sodium--related variables,
respectively, and $l$ (for ``leakage'') terms include all other
ion conductances with slower dynamics. $C$ is the membrane
capacitance.  $V_{\rm K}$ and $V_{\rm Na}$ are ion-specific
reversal potentials, and $V_l$ is defined such that the total
voltage $V$ is exactly zero when the membrane is at rest.
$\bar{g}_{\rm K}$, $\bar{g}_{\rm Na}$ and $\bar{g}_l$ are
empirically determined maximal conductances for the different ion
species,  and the gating
variables $n$, $m$ and $h$ (on the interval $[0,1]$) have their
own voltage dependent dynamics:
\begin{eqnarray}
 dn/dt &=& (0.01V+0.1)(1-n) \exp(-0.1V) - 0.125n\exp(V/80) \nonumber \\
 dm/dt &=& (0.1V+2.5)(1-m) \exp(-0.1V-1.5) - 4m\exp(V/18) \nonumber \\
 dh/dt &=& 0.07(1-h) \exp(0.05V) - h\exp(-0.1V-4).
\end{eqnarray}
We have used the original values for these
parameters, except for changing the signs of the voltages to
correspond to the modern sign convention: $C=1\,\mathrm{\mu
F}/\mathrm{cm}^2$, $\bar{g}_{\rm K}=36\,\mathrm{mS}
/\mathrm{cm}^2$, $\bar{g}_{\rm Na}=120\,\mathrm{mS}
/\mathrm{cm}^2$, $\bar{g}_l=0.3\,\mathrm{mS} /\mathrm{cm}^2$,
$V_{\rm K}=-12\,\mathrm{mV}$, $V_{\rm Na}=+115\,\mathrm{mV}$,
$V_l=+10.613\,\mathrm{mV}$.  We have taken our system to be a $\pi
\times 30^2 \,\mu\mathrm{m}^2$ patch of membrane.
We solve these equations numerically using fourth order
Runge--Kutta integration.

The system is driven with a Gaussian random noise current $I(t)$,
generated by smoothing a Gaussian random number stream with an
exponential filter to generate a correlation time $\tau$. It is
convenient to choose $\tau$ to be longer than the time steps of
numerical integration, since this guarantees that all functions
are smooth on the scale of single time steps.  Here we will always
use $\tau = 0.2$ msec, a value that is both less than the
timescale over which we discretize the stimulus for analysis, and
far less than the neuron's capacitative smoothing timescale $RC
\sim 3$ msec. $I(t)$ has a standard deviation $\sigma$, but since
the correlation time is short the relevant parameter usually is
the spectral density $S = \sigma^2 \tau$; we also add a DC offset
$I_0$. In the following, we will consider two parameter regimes,
$I_0 = 0$, and $I_0$ a finite value, which leads to more periodic
firing.

\begin{figure}[htb]
\begin{center}
\includegraphics[scale=0.65]{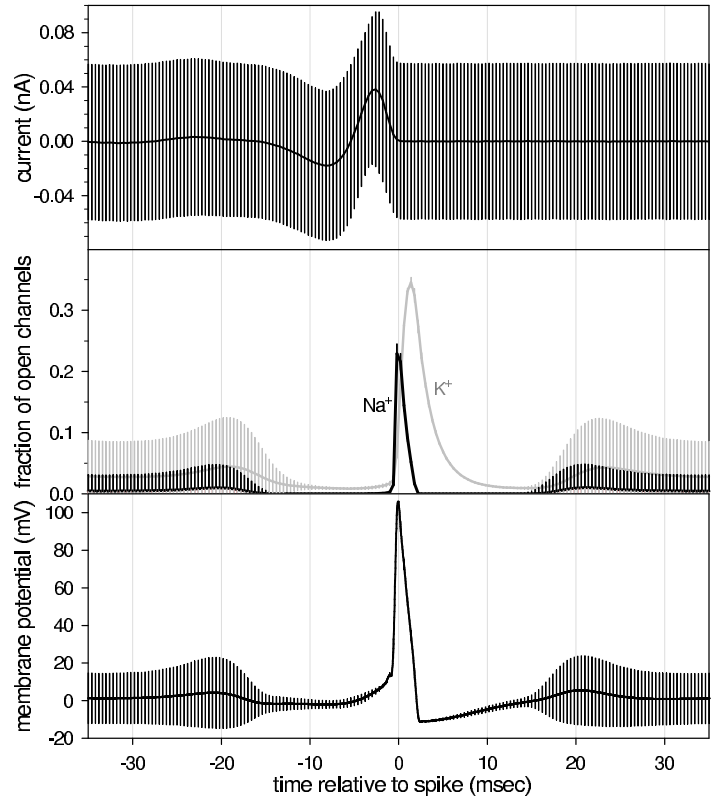}
\end{center}
\caption{Spike triggered averages with standard deviations for (a)
the input current $I$, (b) the fraction of open K$^+$ and Na$^+$
channels, and (c) the membrane voltage $V$, for the parameter
regime $I_0 = 0$ and $S = 6.50 \times 10^{-4} \mathrm{nA}^2
\mathrm{sec}$ .} \label{f:stasville}
\end{figure}

The integration step size is fixed at $0.05\, \mathrm{msec}$. The
key numerical experiments were repeated at a step size of $0.01\,
\mathrm{msec}$ with identical results.  The time of a spike is
defined as the moment of maximum voltage, for voltages exceeding a
threshold (see Fig. \ref{f:stasville}), estimated to subsample
precision by quadratic interpolation. As spikes are both very
stereotyped and very large compared to subspiking fluctuations,
the precise value of this threshold is unimportant; we have used
$+20$ mV.

\subsection{Qualitative description of spiking}

The first step in our analysis is to use reverse correlation,
Eq.~(\ref{e:sta}), to determine the average stimulus feature
preceding a spike, the spike triggered average (STA). In Fig.
\ref{f:stasville}(a) we display the STA in a regime where the
spectral density of the input current is $6.5 \times 10^{-4}\,
\mathrm{nA}^2\mathrm{msec}$. The spike triggered averages of the gating
terms $n^4$ (proportion of open potassium channels) and $m^3h$ (proportion
of open sodium channels) and the membrane voltage $V$ are plotted in parts
(b) and (c).  The errorbars mark the standard deviation of the
trajectories of these variables.

As expected, the voltage and gating variables follow highly
stereotyped trajectories during the $\sim$5 msec surrounding a
spike: First, the rapid opening of the sodium channels causes a
sharp membrane depolarization (or rise in $V$); the slower
potassium channels then open and repolarize the membrane, leaving
it at a slightly lower potential than rest.  The potassium
channels close gradually, but meanwhile the membrane remains
hyperpolarized and, due to its increased permeability to potassium
ions, at lower resistance. These effects make it difficult to
induce a second spike during this $\sim$15 msec ``refractory
period.''  Away from spikes, the resting levels and fluctuations
of the voltage and gating variables are quite small. The larger
values evident in Fig.~\ref{f:stasville}(b) and (c) by $\pm15$
msec are due to the summed contributions of nearby spikes.

The spike triggered average current has a largely transient form,
so that spikes are on average preceded by an upward swing in
current.  On the other hand, there is no obvious ``bottleneck'' in
the current trajectories, so that the current variance is almost
constant throughout the spike.  This is qualitatively consistent
with the idea of dimensionality reduction:  If the neuron ignores
most of the dimensions along which the current can vary, then the
variance---which is shared almost equally among all dimensions for
this near white noise---can change only by a small amount.

\subsection{Interspike interaction}

\label{s:bursty}

Although the STA has the form of a differentiating kernel,
suggesting that the neuron detects edge--like events in the
current vs. time, there must be a DC component to the cell's
response. We recall that for constant inputs the Hodgkin--Huxley
model undergoes a bifurcation to constant frequency spiking, where
the frequency is a function of the value of the input above onset.
Correspondingly the STA does not sum precisely to zero; one might
think of it as having a small integrating component that allows
the system to spike under DC stimulation, albeit only above a
threshold.

\begin{figure}[hbt]
\begin{center}
\includegraphics[scale=0.8]{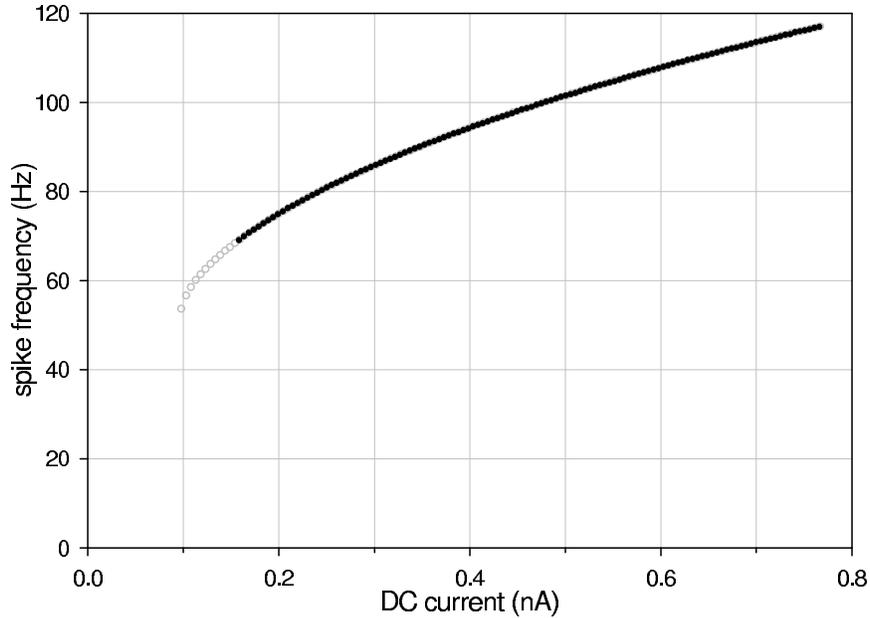}
\end{center}
\caption{Firing rate of the Hodgkin-Huxley neuron as a function of
injected DC current.  The empty circles at moderate currents
denote the metastable region, where the neuron may be either
spiking or silent.} \label{f:fIcurve}
\end{figure}

\begin{figure}[hbt]
\begin{center}
\includegraphics[scale=1.0]{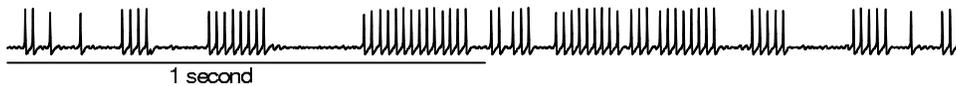}
\end{center}
\caption{Segment of a typical spike train in a ``bursting''
regime.} \label{f:spiketrain}
\end{figure}

The system's tendency to periodic spiking under DC current input
also is felt under dynamic stimulus conditions, and can be thought
of as a strong interaction between successive spikes. We
illustrate this by considering a different parameter regime with a
small DC current and some added noise ($I_0 = 0.11\,\mathrm{nA}$
and $S = 0.8 \times 10^{-4}\, \mathrm{nA}^2 \mathrm{sec}$).  Note
that the DC component puts the neuron in the metastable region of
its $f-I$ curve, Fig.~\ref{f:fIcurve}. In this regime the neuron
tends to fire quasi-regular trains of spikes intermittently, as
shown in Fig. \ref{f:spiketrain}. We will refer to these
quasi--regular spike sequences as ``bursts'' (note that this term
is often used to refer to compound spikes in neurons with
additional channels; such events do not occur in the
Hodgkin--Huxley model).

\begin{figure}[hbt]
\begin{center}
\includegraphics[scale=0.875]{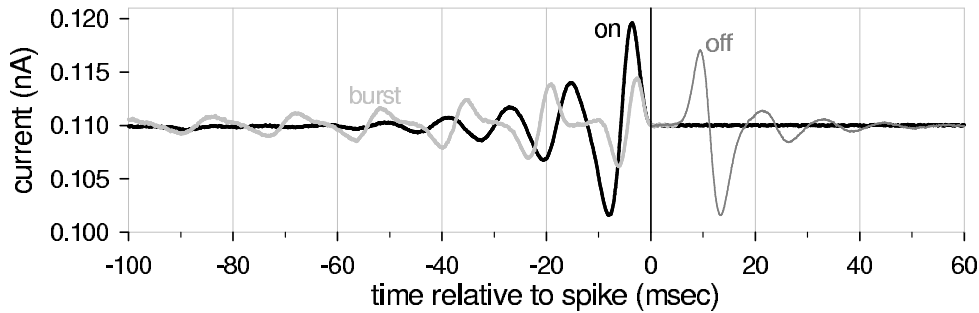}
\end{center}
\caption{Spike triggered averages, derived from spikes leading
(``on''), inside (``burst'') and ending (``off'') a burst. The
parameters of this bursting regime are $I_0 = 0.11\,\mathrm{nA}$
and $S = 0.8 \times 10^{-4}\, \mathrm{nA}^2 \mathrm{sec}$. Note
that the burst-ending spike average is, by construction, identical
to that of any other within-burst spike for $t<0$.}
\label{f:burstSTA}
\end{figure}

Spikes can be classified into three types---those initiating a
spike burst; those within a burst; and those ending a burst. The
minimum length of the silence between bursts is taken in this case
to be $70\, \mathrm{msec}$.  Taking these three categories of
spike as different ``symbols'' \cite{bialek88}, we can determine the
average stimulus for each. These are shown in Fig. \ref{f:burstSTA} with
the spike at $t = 0$.

In this regime, the initial spike of a burst is preceded by a
rapid oscillation in the current.  Spikes within a burst are
affected much less by the current; the feature immediately
preceding such spikes is similar in shape to a single
``wavelength'' of the leading spike feature, but is of much
smaller amplitude, and is temporally compressed into the
interspike interval. Hence, although it is clear that the timing
of a spike within a burst is determined largely by the timing of
the previous spike, the current plays some role in affecting the
precise placement. This also demonstrates that the shape of the
STA is not the same for all spikes; it depends strongly and
nontrivially on the time to the previous spike, and this is
related to the observation that subtly different patterns of two
or three spikes correspond to very different average stimuli
\cite{bialek88}. For a reader of the spike code, a spike within a
burst conveys a different message about the input than the spike
at the onset of the burst. Finally, the feature ending a burst has
a very similar form to the onset feature, but reversed in time.
Thus, to a good approximation, the {\em absence} of a spike at the
end of a burst can be read as the opposite of the onset of the
burst.

\begin{figure}[htb]
\begin{center}
\includegraphics[scale=0.8]{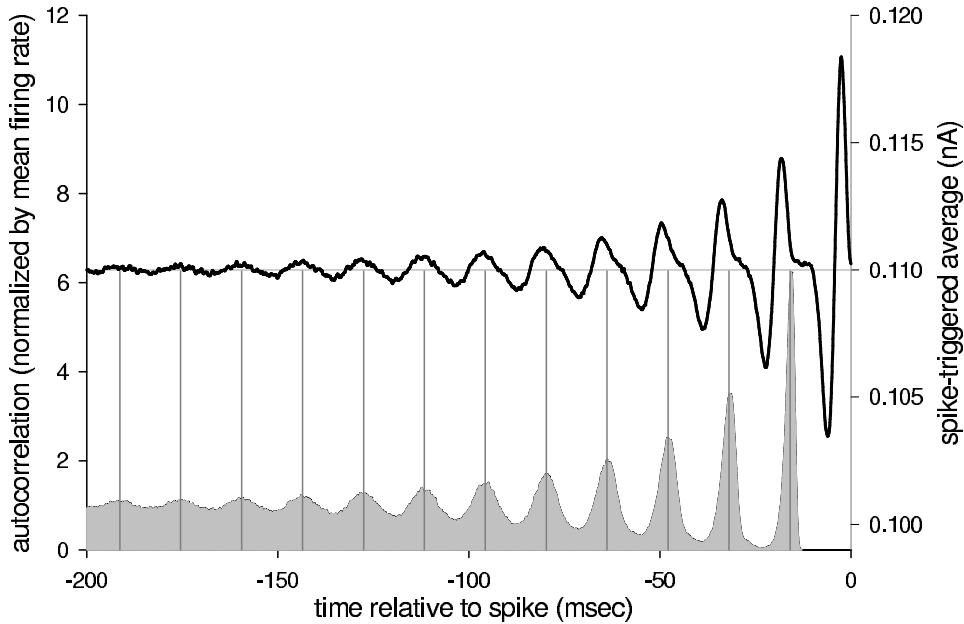}
\end{center}
\caption{Overall spike triggered average in the bursty regime,
showing the ringing due to the tendency to periodic firing;
plotted in grey is the spike autocorrelation, showing the same
oscillations.} \label{f:allspksSTA}
\end{figure}

In summary, this regime of the HH neuron is similar to a
``flip-flop'', or 1-bit memory.  Like its electronic analogue, the
neuron's memory is preserved by a feedback loop, here implemented
by the interspike interaction.  Large fluctuations in the input
current at a certain frequency ``flip'' or ``flop'' the neuron
between its silent and spiking states. However, while the neuron
is spiking, further details of the input signal are transmitted by
precise spike timing within a burst.  If we calculate the spike
triggered average of all spikes for this regime, without regard to
their position within a burst, then as shown in Fig. \ref{f:allspksSTA}
the relatively well localized leading spike oscillation of
Fig.~\ref{f:burstSTA} is replaced by a long-lived oscillating
function resulting from the spike periodicity.  This is shown
explicitly by comparing the overall STA with the spike
autocorrelation, also shown in Fig.~\ref{f:allspksSTA}.  This same
effect is seen in the STA of the burst spikes, which in fact
dominates the overall average. Prediction of spike timing using
such an STA would be computationally difficult, due to its
extension in time, but, more seriously, unsuccessful, as most of
the function is an artifact of the spike history rather than the
effect of the stimulus.

\begin{figure}[htb]
\begin{center}
\includegraphics[scale=0.8]{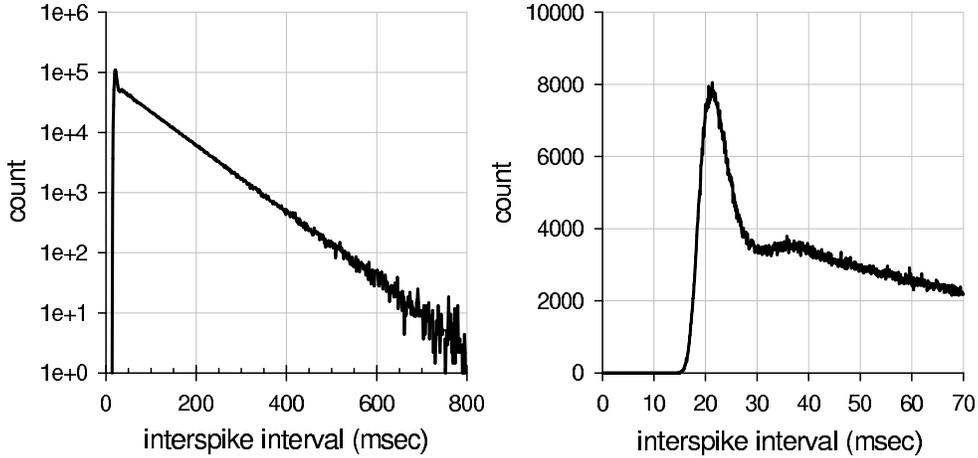}
\end{center}
\caption{Hodgkin-Huxley interspike interval histogram for the
parameters $I_0 = 0$ and $S = 6.5 \times 10^{-4}\, \mathrm{nA}^2
\mathrm{sec}$, showing a peak at a preferred firing frequency and
the long Poisson tail. The total number of spikes is $N = 5.18
\times 10^6$. The plot to the right is a closeup in linear scale.}
\label{f:isid}
\end{figure}

While the effects of spike interaction are interesting, and should
be included in a complete model for spike generation, we wish here
to consider only the current's role in initiating spikes.
Therefore, as we have argued elsewhere, we limit ourselves
initially to the cases in which interspike interaction plays no
role \cite{nips2000,i&f}. These ``isolated'' spikes can be defined
as spikes preceded by a silent period $t_{\rm silence}$ long
enough to ensure decoupling from the timing of the previous spike.
A reasonable choice for $t_{\rm silence}$ can be inferred directly
from the interspike interval distribution $P(\Delta t)$,
illustrated in Fig.~\ref{f:isid}. For the HH model, as in simpler
models and many real neurons \cite{naamaISID}, the form of
$P(\Delta t)$ has three noteworthy features: a refractory ``hole''
during which another spike is unlikely to occur, a strong mode at
the preferred firing frequency, and an exponentially decaying, or
Poisson tail. The details of all three of these features are
functions of the parameters of the stimulus, and certain regimes
may be dominated by only one or two features. The emergence of
Poisson statistics in the tail of the distribution implies that
these events are independent, so we can infer that the system has
lost memory of the previous spike. We will therefore take isolated
spikes to be those preceded by a silent interval $\Delta t \ge
t_{\rm silence}$, where $t_{\rm silence}$ is well into the Poisson
regime. Note that the burst onset spikes of Fig.~\ref{f:burstSTA}
are isolated spikes by this definition.

\section{Isolated spike analysis}

\begin{figure}[htb]
\begin{center}
\includegraphics[scale=0.85]{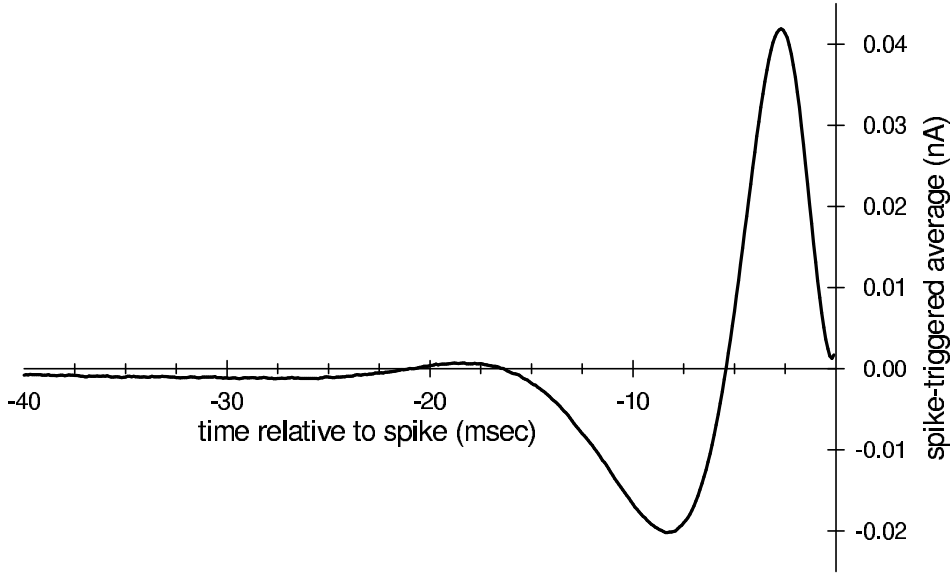}
\end{center}
\caption{Spike triggered average stimulus for isolated spikes.}
\label{f:isosta}
\end{figure}

Focusing now on isolated spikes, we proceed to a second--order
analysis of the current fluctuations around the isolated spike
triggered average, Fig.~\ref{f:isosta}. We consider the response
of the HH neuron to currents $I(t)$ with mean $I_0 = 0$ and
spectral density of $S = 6.5 \times 10^{-4}\, \mathrm{nA}^2
\mathrm{sec}$. Isolated spikes in this regime are defined by
$t_{\rm silence} = 60$ msec.

\subsection{How many dimensions?}
\label{s:howmanydim}

\begin{figure}[htb]
\begin{center}
\flushleft
\includegraphics[scale=0.85]{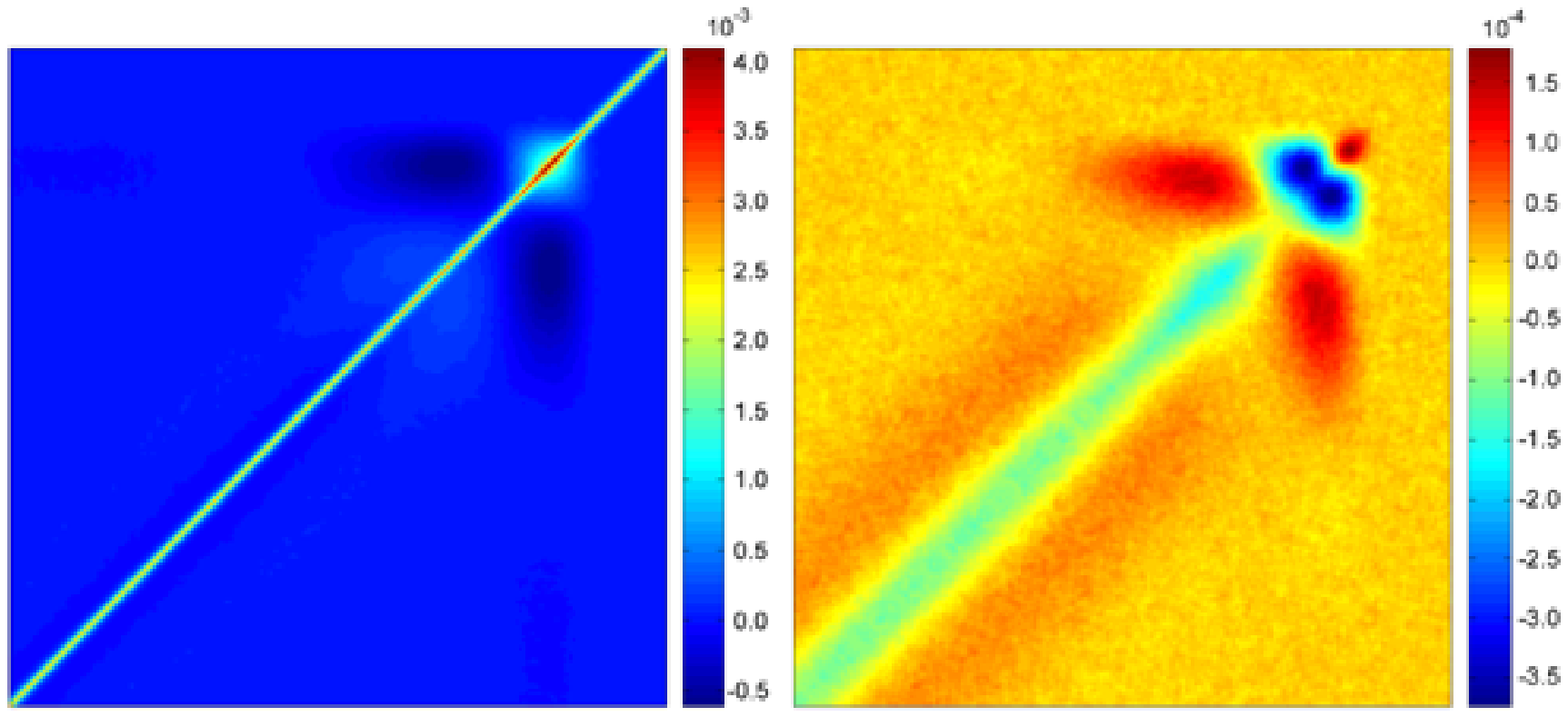}
\end{center}
\caption{The isolated spike triggered covariance $C_{\rm iso}$
(left) and covariance difference $\Delta C$ (right) for times $-30
< t < 5$ msec. The plots are in units of $\mathrm{nA}^2$.}
\label{f:cov-matrices}
\end{figure}

As explained in Section 2, our path to dimensionality reduction
begins with the computation of covariance matrices for stimulus
fluctuations surrounding a spike.  The matrices are accumulated
from stimulus segments 200 samples in length, roughly
corresponding to sampling at the correlation timescale for
sufficiently long to capture the relevant features. Thus we begin
in a 200 dimensional space. We emphasize that the theorem which
connects eigenvalues of the matrix $\Delta C$ to the number of
relevant dimensions is valid only for truly Gaussian distributions
of inputs, and that by focusing on isolated spikes we are
essentially creating a nonGaussian stimulus ensemble, namely those
stimuli which generate the silence out of which the isolated spike
can appear.  Thus we expect that the covariance matrix approach
will give us a heuristic guide to our search for lower dimensional
descriptions, but we should proceed with caution.

\begin{figure}[htb]
\begin{center}
\includegraphics[scale=0.9]{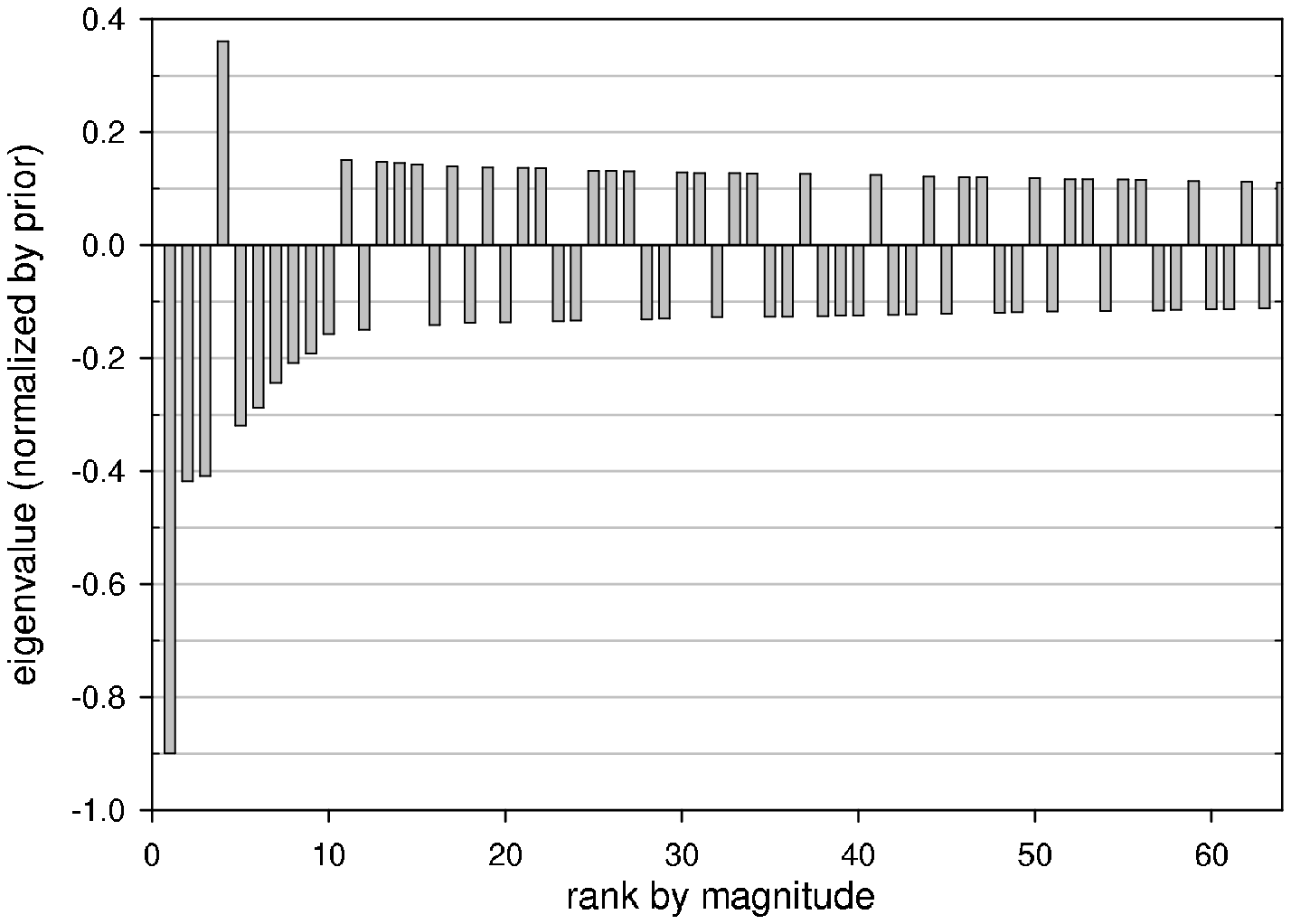}
\end{center}
\caption{The leading 64 eigenvalues of the isolated spike
triggered covariance, after accumulating 80,000 spikes.}
\label{f:cov-eig80k}
\end{figure}

The ``raw'' isolated spike triggered covariance
$C_{\rm iso\, spike}$ and the corresponding covariance difference $\Delta C$,
Eq.~(\ref{e:deltac}), are shown in Fig.~\ref{f:cov-matrices}.  The matrix
shows the effect of the silence as an approximately translationally
invariant band preceding the spike, the second order analogue of the
constant negative bias in the isolated spike STA, Fig.\ref{f:isosta}.
The spike itself is associated with features localized to $\pm
15\,$msec.
In Fig.~\ref{f:cov-eig80k} we show the spectrum of eigenvalues
of $\Delta C$ computed using a sample of 80,000
spikes. Before calculating the spectrum, we multiply
$\Delta C$ by $C_{\rm prior}^{-1}$.  This has the effect of giving
us eigenvalues scaled in units of the input standard deviation
along each dimension. Because the correlation time is short,
$C_{\rm prior}$ is nearly diagonal.

\begin{figure}[htb]
\begin{center}
\includegraphics[scale=0.9]{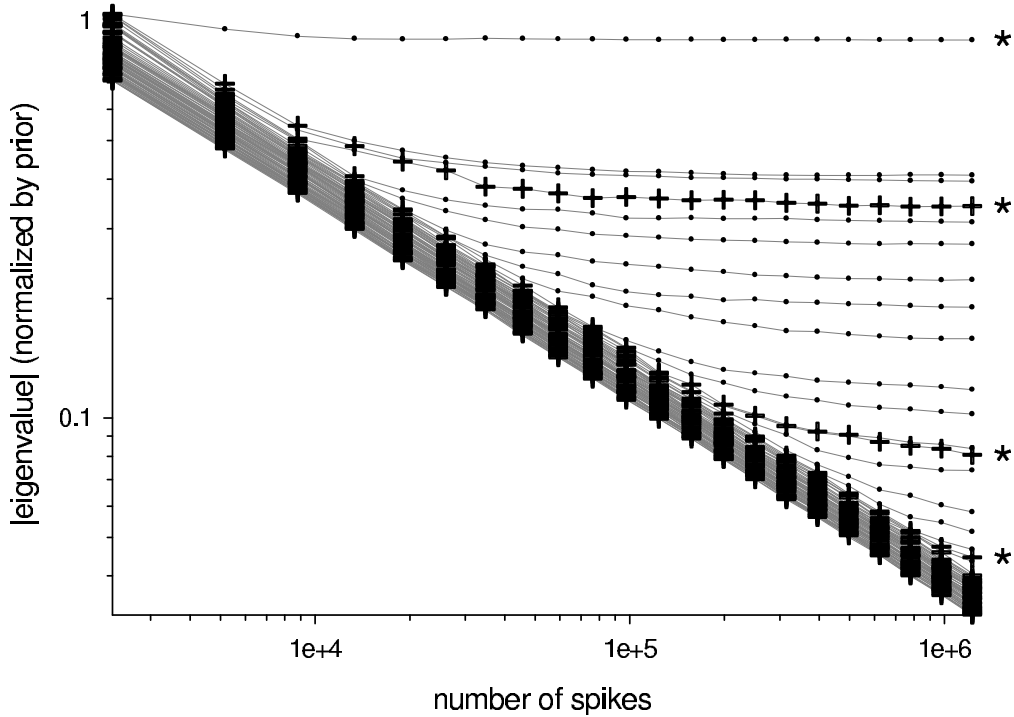}
\end{center}
\caption{Convergence of the leading 64 eigenvalues of the isolated
spike triggered covariance with increasing sample size.  The log
slope of the diagonal is $1/\sqrt{n_{\mathrm{spikes}}}$. Positive
eigenvalues are indicated by crosses and negative by dots. The
spike-associated modes are labelled with an asterisk.}
\label{f:cov-convergence}
\end{figure}

While the eigenvalues decay rapidly,  there is no obvious set of
outstanding eigenvalues.  To verify that this is not an effect of finite
sampling, Fig.~\ref{f:cov-convergence} shows the spectrum of eigenvalue
magnitudes as a function of sample size $N$. Eigenvalues which are
truly zero up to the noise floor determined by sampling decrease
like $\sqrt{N}$.  We find that a sequence of eigenvalues emerges
stably from the noise.

\begin{figure}[htb]
\begin{center}
\includegraphics[scale=0.8]{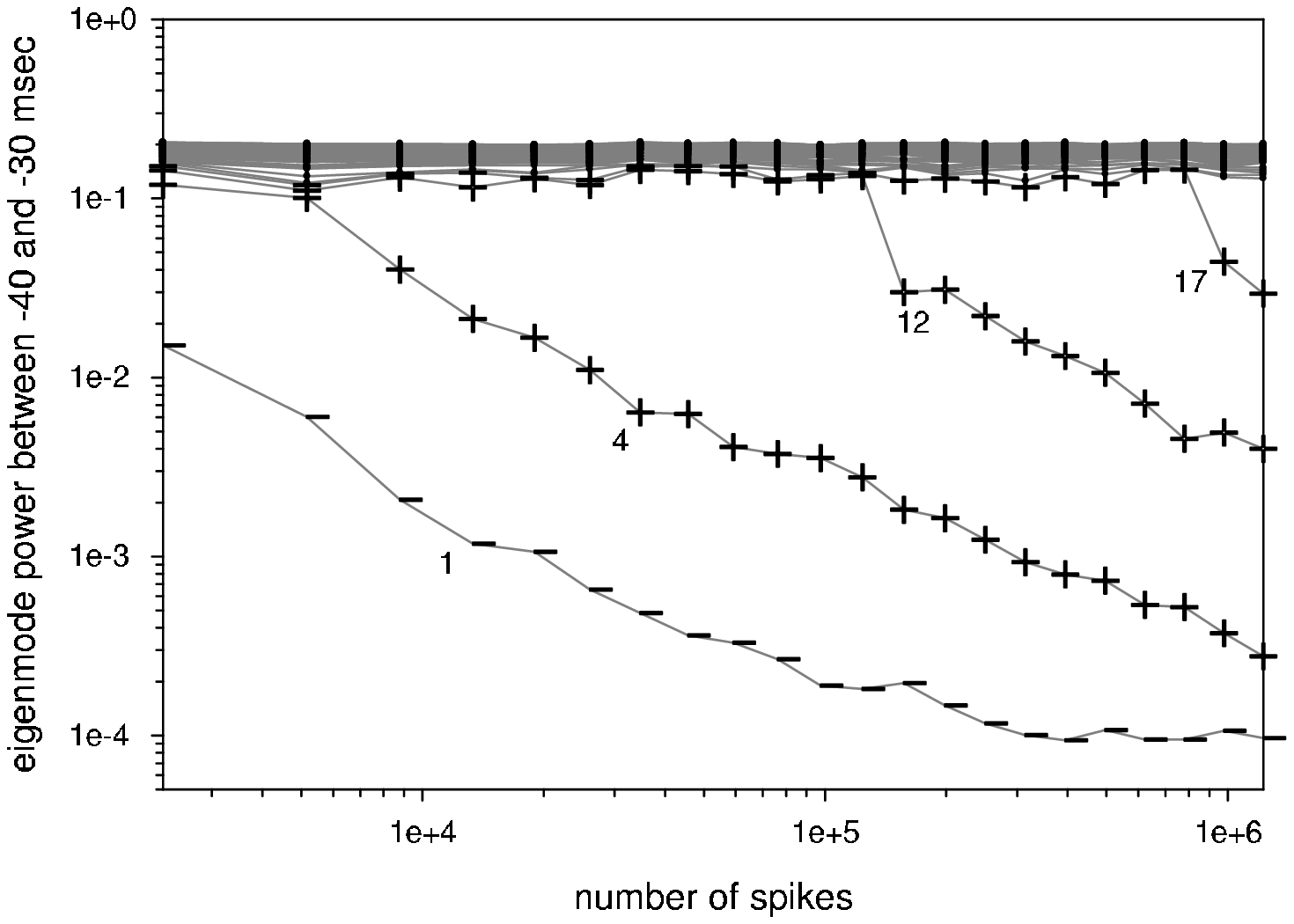}
\end{center}
\caption{For the leading 64 modes, fraction of the mode energy
over the interval $-40 <t<-30$ msec as a function of increasing
sample size.  Modes emerging with low energy are
spike--associated. The symbols indicate the sign of the
eigenvalue.} \label{f:silencePower}
\end{figure}

\begin{figure}[htb]
\begin{center}
\includegraphics[scale=0.9]{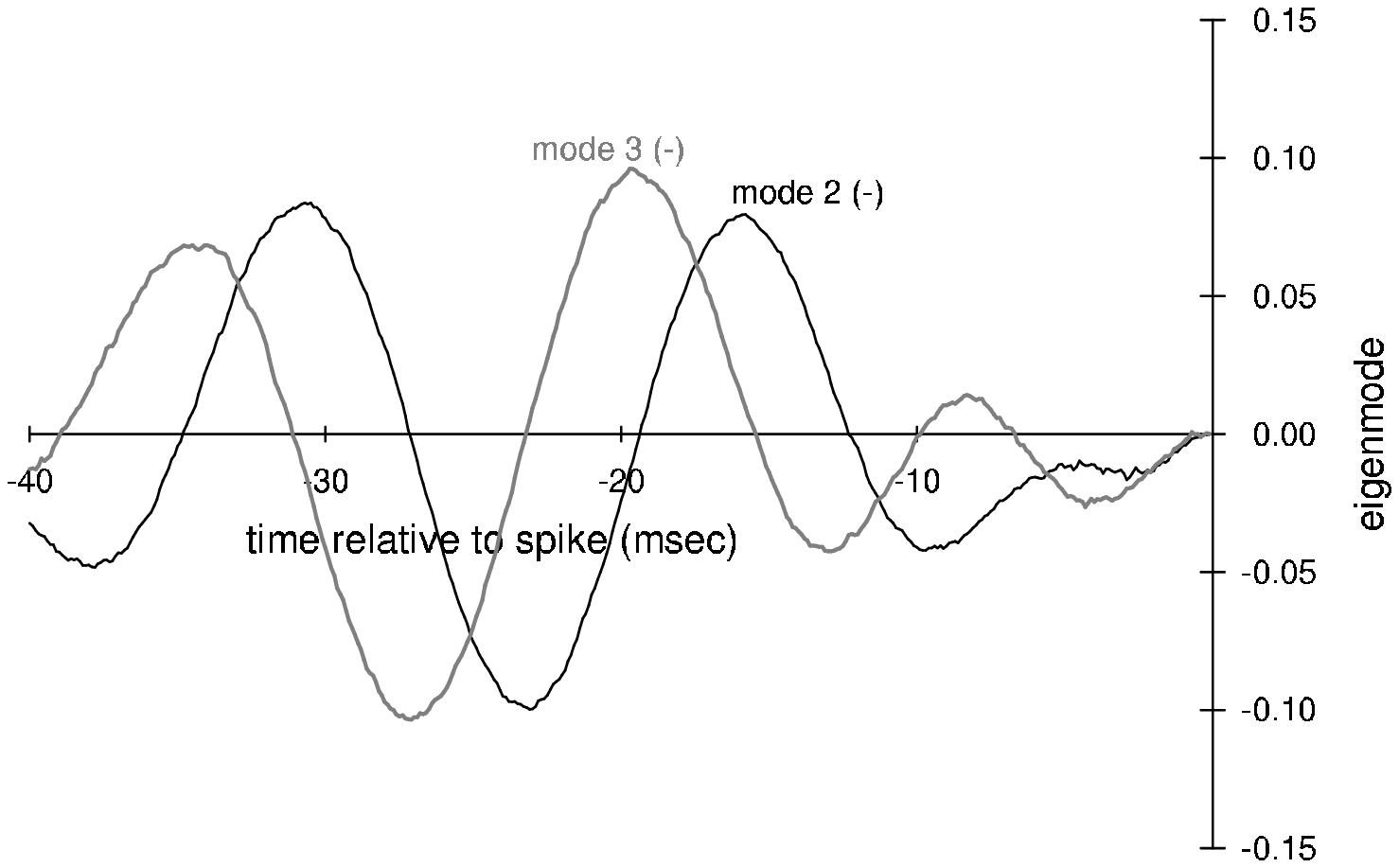}
\end{center}
\caption{Modes 2 and 3 of the spike triggered covariance
(silence-associated).} \label{f:silence-modes}
\end{figure}

These results do not, however, imply that a low dimensional
approximation cannot be identified.   The extended structure in
the covariance matrix induced by the silence requirement is
responsible for the apparent high dimensionality. In fact, as has
been shown in \cite{i&f}, the covariance eigensystem includes
modes that are local and spike--associated, and others that are
extended and silence-associated, and thus irrelevant to a causal
model of spike timing prediction. Fortunately, because extended
silences and spikes are (by definition) statistically independent,
there is no mixing between the two types of modes. To identify the
spike--associated modes, we follow the diagnostic of \cite{i&f},
computing the fraction of the energy of each mode concentrated in
the period of silence, which we take to be $-60 \leq t \leq -40$
msec.  The energy of a spike--associated mode in the silent period
is due entirely to noise, and will therefore decrease like
$1/n_{\rm spikes}$ with increasing sample size, while this energy
remains of order unity for silence modes. Carrying out the test on
the covariance modes, we obtain Fig.~\ref{f:silencePower}, which
shows that the first and fourth modes rapidly emerge as
spike--associated. Two further spike--associated modes appear over
the sample shown, with the suggestion of other, weaker modes yet
to emerge.  The two leading silence modes are shown in
Fig.~\ref{f:silence-modes}. Those shown are typical; most modes
resemble Fourier modes, as the silence condition is close to time
translationally invariant.

\begin{figure}[htb]
\begin{center}
\includegraphics[scale=0.9]{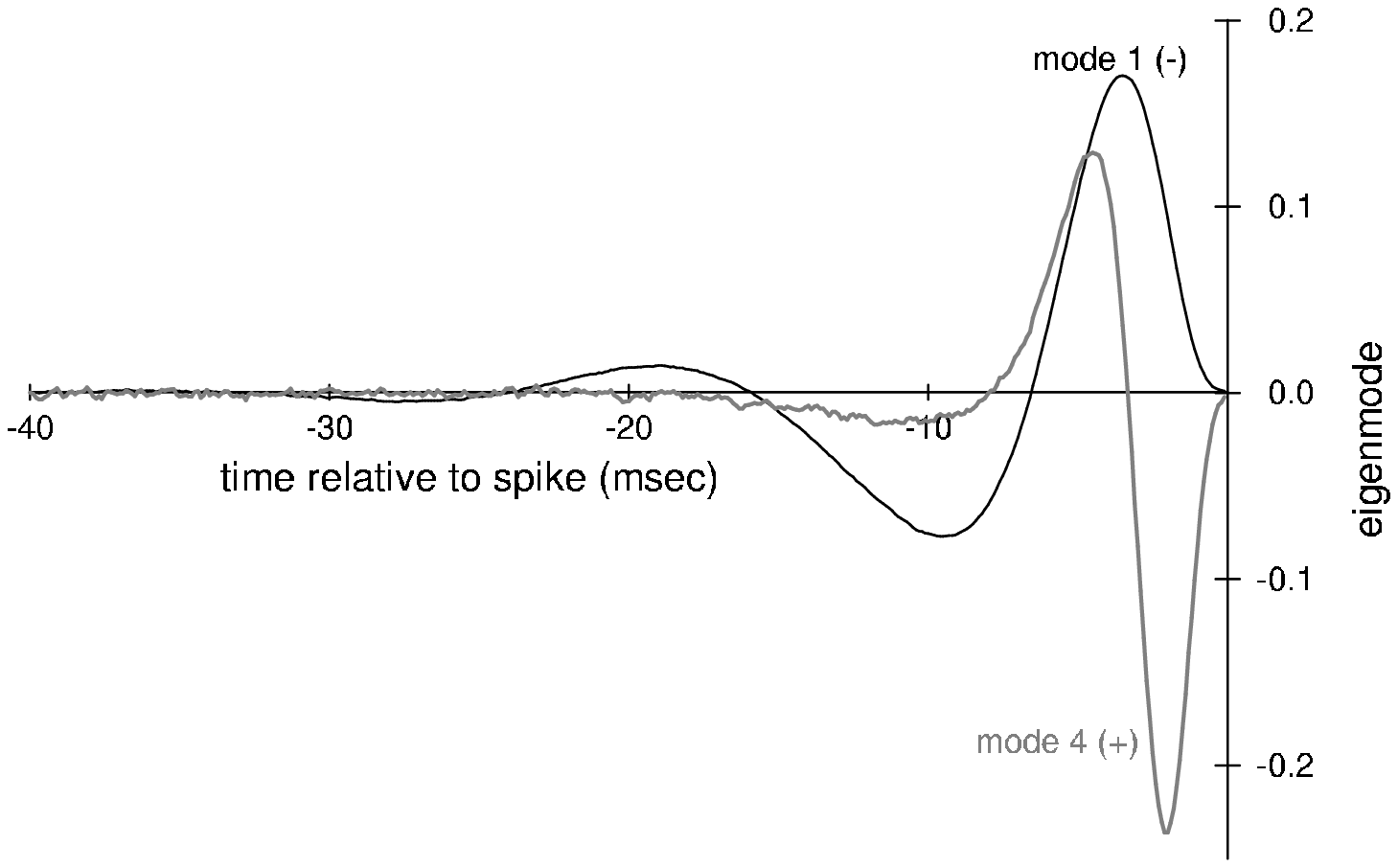}
\end{center}
\caption{Modes 1 and 4 of the spike triggered covariance, which are the
leading spike--associated modes.} \label{f:spike-modes}
\end{figure}

Examining the eigenvectors corresponding to the two leading
spike--associated eigenvalues, which for convenience we will denote
$s_1$ and $s_2$ (although they are not the leading modes of the
matrix), we find (Fig.~\ref{f:spike-modes}) that the first mode
closely resembles the isolated spike STA, and the second is close
to the derivative of the first.  Both modes approximate
differentiating operators; there is no linear combination of these
modes which would produce an integrator.

If the neuron filtered its input and generated a spike when the
output of the filter crosses threshold, we would find two
significant dimensions associated with a spike. The first
dimension would correspond simply to the filter, as the variance
in this dimension is reduced to zero (for a noiseless system) at
the occurrence of a spike. As the threshold is always crossed from
below, the stimulus projection onto the filter's {\it derivative}
must be positive, again resulting in a reduced variance. It is tempting to
suggest, then, that filtered threshold crossing is a good approximation
to the HH model, but we will see that this is not correct.

\subsection{Evaluating the nonlinearity}

\begin{figure}[htb]
\begin{center}
\includegraphics[scale=3]{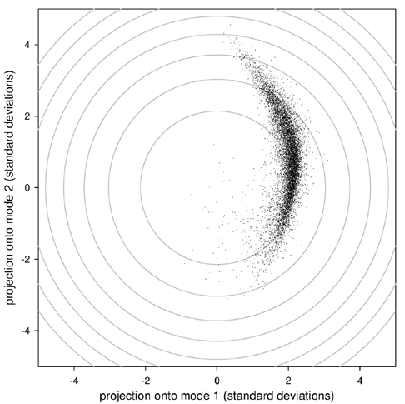}
\end{center}
\caption{$10^4$ spike conditional stimuli (or ``spike histories'')
projected along the first two covariance modes. The axes are in
units of standard deviation on the prior Gaussian distribution.
The circles, from the inside out, enclose all but $10^{-1},
10^{-2}, \dots , 10^{-8}$ of the prior.} \label{f:cov-scatter}
\end{figure}

At each instant of time we can find the projections of the
stimulus along the leading spike--associated dimensions $s_1$ and
$s_2$. By construction, the distribution of these signals over the
whole experiment, $P(s_1, s_2)$, is Gaussian. The appropriate
prior for the isolation condition, $P(s_1, s_2|{\rm silence})$,
differs only subtly from the Gaussian prior.  On the other hand,
for each spike we obtain a sample from the distribution $P(s_1,
s_2 | {\rm iso~ spike \,at \, }t_0)$, leading to the picture in
Fig.~\,\ref{f:cov-scatter}. The prior and spike conditional
distributions are clearly better separated in two dimensions than
in one, which means that the two dimensional description captures
more information than projection onto the spike triggered average
alone. Surprisingly, the spike conditional distribution is curved,
unlike what we would expect for a simple thresholding device.
Furthermore, the eigenvalue of $\Delta C$ which we associate with
the direction of threshold crossing (plotted on the $y$-axis in
Fig.~\,\ref{f:cov-scatter}) is {\em positive}, indicating
increased rather than decreased variance in this direction.  As we
see, projections onto this mode are almost equally likely to be
positive or negative, ruling out the threshold crossing
interpretation.

Combining eqs. (\ref{e:Kprojs}) and (\ref{e:Bayes}) for isolated
spikes, we have
\begin{equation}
g(s_1 , s_2 ) = {{P(s_1, s_2 | \,{\rm iso\, spike \, at \,}t_0
)}\over{P(s_1, s_2|\,{\rm silence} )}} ,
\end{equation}
so that these two distributions determine the input/output
relation of the neuron in this 2D space \cite{naama}.  Recall that
although the subspace is linear, $g$ can have arbitrary
nonlinearity. Fig.~\ref{f:cov-scatter} shows that this
input/output relation has clear structure, but also some
fuzziness. As the HH model is deterministic, the input/output
relation should be a singular function in the continuous space of
inputs---spikes occur only when certain exact conditions are met.
Of course, finite time resolution introduces some blurring, and so
we need to understand whether  the blurring of the input/output
relation in Fig.~\ref{f:cov-scatter} is an effect of finite time
resolution or a real limitation of the 2D description.

\subsection{Information captured in two dimensions}

\begin{figure}[htb]
\begin{center}
\includegraphics[scale=0.75]{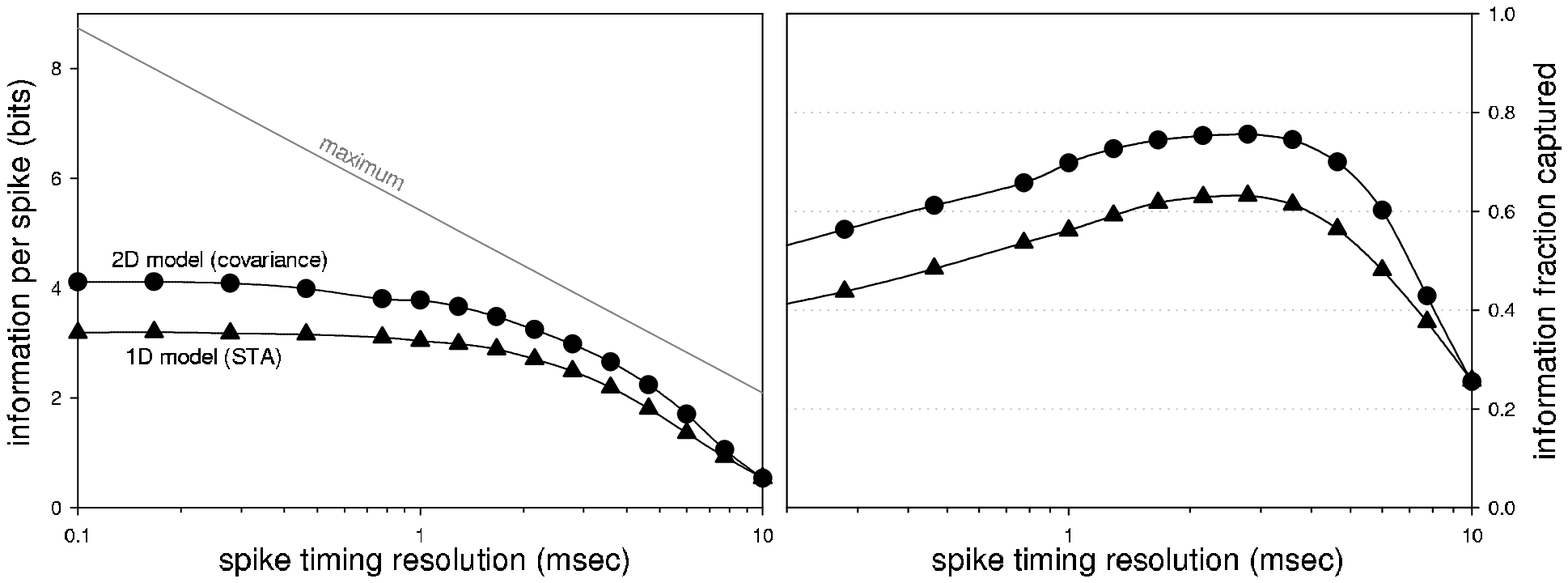}
\end{center}
\caption{Bits per spike (left) and fraction of the theoretical
limit (right) of timing information in a single spike at a given
temporal resolution captured by projection onto the STA alone
(triangles), and projection onto $\Delta C$ covariance modes 1 and
2 (circles).} \label{f:info-fraction}
\end{figure}

We will measure the effectiveness of our description by computing
the information in the 2D approximation, according to the methods
described in Sect.~\ref{s:info}.
If the two dimensional approximation were exact we would find that
$I_{\rm iso\, spike}^{s_1,s_2} = I_{\rm iso\, spike}$; more
generally one finds $I_{\rm iso\, spike}^{s_1,s_2} \leq I_{\rm
iso\, spike}$, and the fraction of the information captured
measures the quality of the approximation.  This fraction is
plotted in Fig.~\ref{f:info-fraction} as a function of time
resolution.  For comparison, we also show the information captured
in the one dimensional case,  considering only the stimulus
projection along the STA.

We find that our low dimensional model captures a substantial
fraction of the total information available in spike timing in a
HH neuron over a range of time resolutions.  The approximation is
best near $\Delta t = 3 \,\mathrm{msec}$, reaching 75\%.  Thus the
complex nonlinear dynamics of the Hodgkin--Huxley model can be
approximated by saying that the neuron is sensitive to a two
dimensional linear subspace in the high dimensional space of input
signals, and this approximate description captures up to 75\% of
the mutual information between input currents and (isolated) spike
arrival times.

The dependence of information on time resolution, Fig.
\ref{f:info-fraction}, shows that the absolute information
captured saturates for both the 1D and 2D cases, at $\approx 3.2$
and $4.1$ bits respectively. Hence, for smaller $\Delta t$, the
information fraction captured drops. The model provides, at its
best, a time resolution of 3 msec, so that information carried by
more precise spike timing is lost in our low dimensional
projection. Might this missing information be important for a real
neuron? Stochastic HH simulations with realistic channel densities
suggest that the timing of spikes in response to white noise
stimuli is reproducible to within 1--2 msec \cite{elad}, a figure
which is comparable to what is observed for pyramidal cells in
vitro \cite{mainensejnowski}, as well in vivo in the fly's visual system
\cite{billscience,flyNaturalStim}, the vertebrate retina
\cite{berryWarlandMeister}, the cat LGN \cite{reinagel&reid} and
the bat auditory cortex \cite{dear&simmons}. This suggests that
such timing details may indeed be important.  We must therefore
ask why our approximation seems to carry an inherent time
resolution limitation, and why, even at its optimal resolution,
the full information in the spike is not recovered.

For many purposes, recovering 75\% of the information at $\sim
3\,$ msec resolution might be considered a resounding success.  On
the other hand, with such a simple underlying model we would hope
for a more compelling conclusion.  From a methodological point of
view it behooves us to ask what we are missing in our 2D model,
and perhaps the methods we use in finding the missing information
in the present case will prove applicable more generally.

\section{What is missing?}

\begin{figure}[htb]
\begin{center}
\includegraphics[scale=0.9]{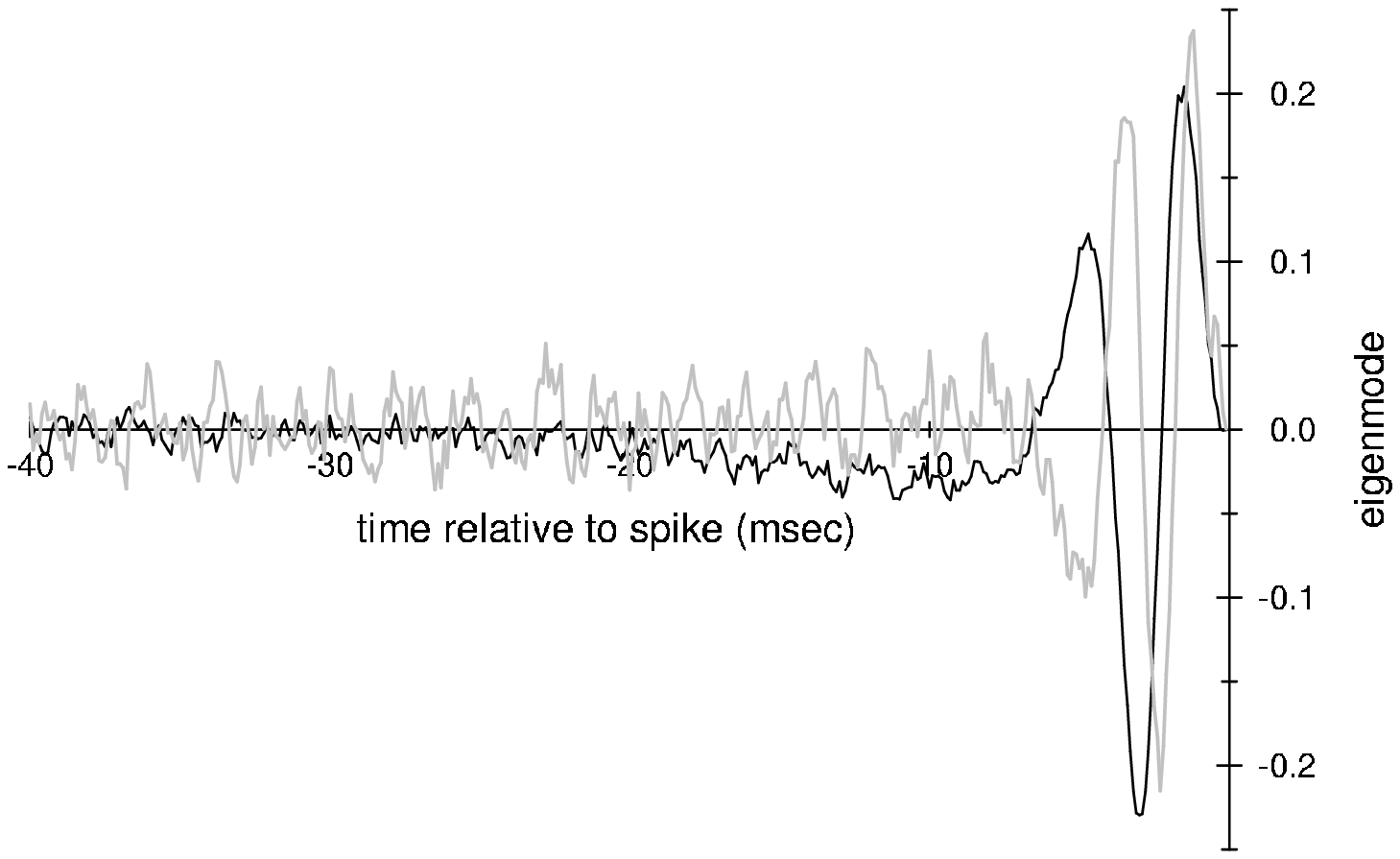}
\end{center}
\caption{The two next spike--associated modes. These resemble
higher-order derivatives.} \label{f:othermodes}
\end{figure}

The obvious first approach to improving the 2D approximation is to
add more dimensions.  Let us consider the neglected modes.  We
recall from Fig.~\ref{f:silencePower} that, in simulations with
very large numbers of spikes, we can isolate at least two more
modes that have significant eigenvalues and are associated with
the isolated spike rather than the preceding silence; these are
shown in Fig.~\ref{f:othermodes}. We see that these modes look
like higher order derivatives, which makes some sense since we are
missing information at high time resolution.  On the other hand,
if all we are doing is expanding in a basis of increasingly higher
order derivatives it is not clear that we will do qualitatively
better by including one or two more terms.

Our original model attempted to approximate the conditional
response ensemble as lying within a $K$--dimensional linear
subspace of the original $D$--dimensional input space. The obvious
generalization of this idea is to regard the subspace as a curved, but
still low dimensional manifold.  Several methods have been proposed for
the general problem of identifying low dimensional nonlinear manifolds
\cite{nonlinearPCA2,svm1,svm2,nonlinearPCA1,roweis}, but these various
approaches share the disadvantage that the manifold or,
equivalently, the relevant set of features remains implicit. Our
hope is to understand the behaviour of the neuron explicitly; we
therefore wish to obtain an explicit representation of the spiking
surface.

\begin{figure}[hbt]
\begin{center}
\includegraphics[scale=0.87]{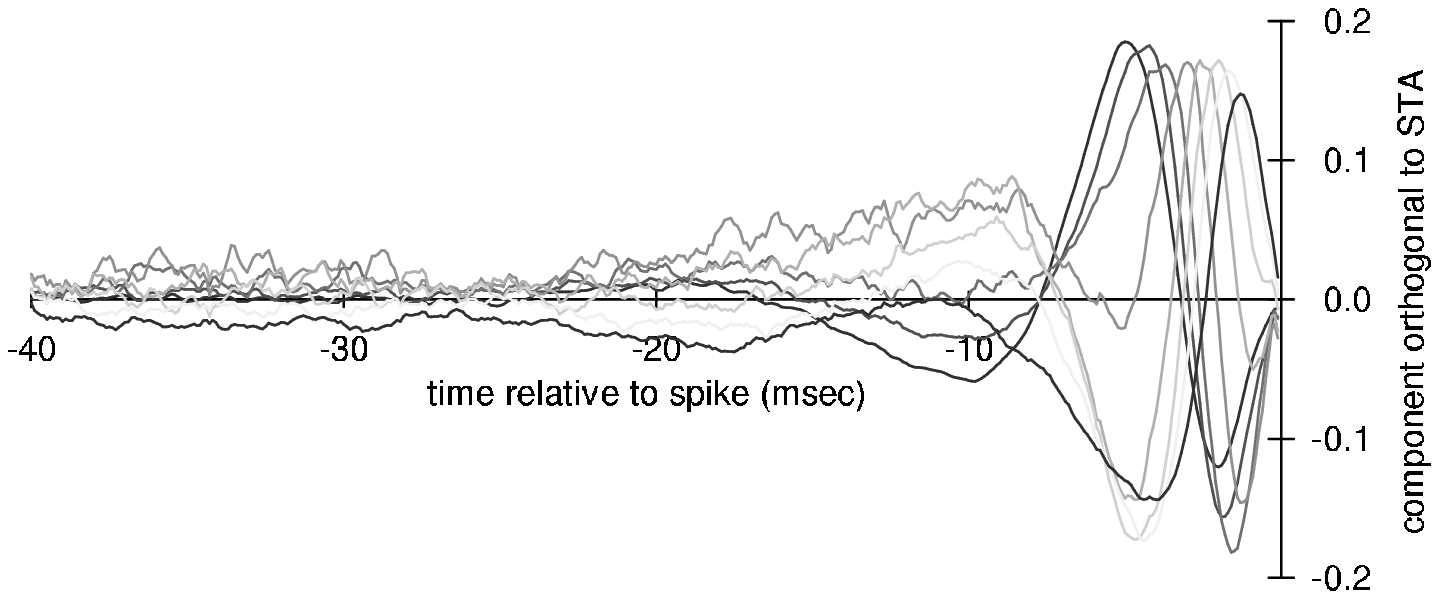}
\end{center}
\caption{The orthonormal components of spike triggered averages
from 80,000 spikes conditioned on their projection onto the
overall spike triggered average (8 conditional averages shown).}
\label{f:condSTAsta}
\end{figure}

A first approach to representing the curved manifold is to try to
form a locally linear tiling.    Beginning
with first order statistics, there is only one natural direction
along which to parametrize---the spike triggered average. We sort
the stimulus histories according to their projection onto the
spike triggered average, divide them into bins, and recompute the
averages over the bins individually.  If these conditional
averages have a component orthogonal to the overall STA, then this
component is a locally meaningful basis vector for the manifold at
that ``slice.''. This procedure results in the family of curves
orthogonal to the STA shown in Fig.~\ref{f:condSTAsta}.

The analysis shown here was carried out with  80,000 isolated
spikes; note that a similar number of spikes cannot resolve more than two
spike--associated covariance modes in the covariance matrix analysis.  The
spikes were divided into eight bins according to their projection onto
the STA, with 10,000 spikes per bin. Applying singular value
decomposition to the family of curves shows that there are at least four
significant independent directions in stimulus space apart from the STA.
This gives us a lower bound on the embedding dimension of the manifold.

\begin{figure}[hbt]
\begin{center}
\includegraphics[scale=0.75]{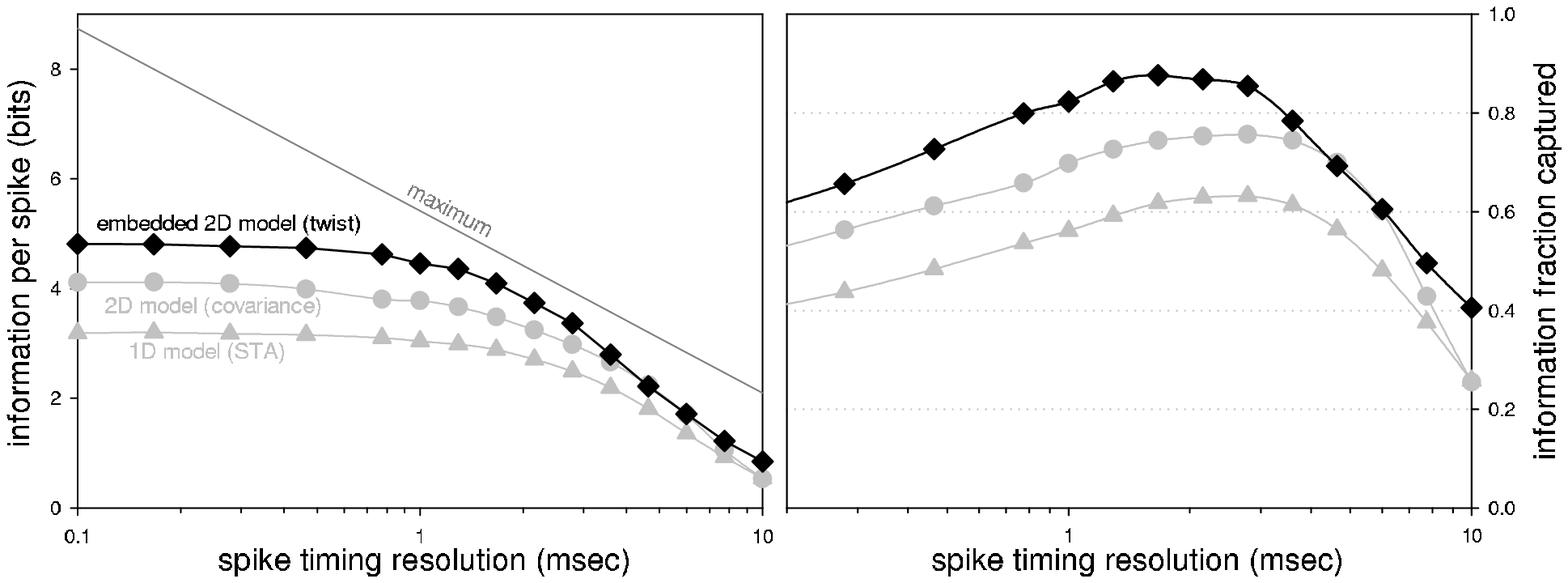}
\end{center}
\caption{Bits per spike (left) and fraction of the theoretical
limit (right) of timing information in a single spike at a given
temporal resolution captured by the locally linear tiling `twist'
model (diamonds), compared to models using the STA alone
(triangles), and projection onto $\Delta C$ covariance modes 1 and
2 (circles).} \label{f:twist-frac}
\end{figure}

Computing the information as a function of $\Delta t$ using this
locally linear model, we obtain the curve shown in
Fig.\ref{f:twist-frac}, where the results can be compared against
the information found from the STA alone and from the covariance
modes. The information from the new model captures a maximum of
4.8 bits, recovering $\sim 90\%$ of the information at a  time
resolution of approximately 1 msec.

\section{Discussion}

The Hodgkin--Huxley equations describe the dynamics of four
degrees of freedom, and almost since these equations first were
written down there have been attempts to find simplifications or
reductions.  FitzHugh and Nagumo proposed a two dimensional system
of equations that approximate the HH model \cite{fitzhugh,nagumo},
and this has the advantage that one can visualize the trajectories
directly in the plane and thus achieve an intuitive graphical
understanding of the dynamics and its dependence on parameters.
The need for reduction in the sense pioneered by FitzHugh and by
Nagumo et al. has become only more urgent with the growing use of
increasingly complex HH--style model neurons with many different
channel types. With this problem in mind, Kepler, Abbott and
Marder have introduced reduction methods which are more
systematic, making use of the difference in timescales among the
gating variables \cite{kepler,abbott}.

In the presence of constant current inputs, it makes sense to
describe the Hodgkin--Huxley equations as a four dimensional
autonomous dynamical system; by well known methods in dynamical
systems theory, one could consider periodic input currents by
adding an extra dimension.  The question asked by FitzHugh and
Nagumo was whether this four or five dimensional description could
be reduced to two or three dimensions.

Closer in spirit to our approach is the work by Kistler et al.,
who focused in particular on the interaction among successive
action potentials \cite{gerstner}.  They argued that one could
approximate the Hodgkin--Huxley model by a nearly linear dynamical
system with a threshold, identifying threshold crossing with
spiking, provided that each spike generated either a change in
threshold or an effective input current which influences the
generation of subsequent spikes.

The notion of model dimensionality considered here is distinct
from the dynamical systems perspective in which one simply counts
the system's degrees of freedom.  Here we are attempting to find a
description of the dynamics which is essentially functional or
computational. We have identified the output of the system as
spike times, and our aim is to construct as complete a description
as possible of the mapping between input and output. The
dimensionality of our model is that of the space of inputs
relevant for this mapping.  There is no necessary relationship
between these two notions of dimensionality.  For example, in a
neural network with two attractors, a system described by a
potentially large number of variables, there might be a simple
rule (perhaps even a linear filter) which allows us to look at the
inputs to the network and determine the times at which the
switching events will occur.  Conversely, once we leave the
simplified world of constant or periodic inputs, even the small
number of differential equations describing a neuron's channel
dynamics could in principle be equivalent to a very complicated
set of rules for mapping inputs into spike times.

In our context, simplicity is (roughly) feature selectivity: the
mapping is simple if spiking is determined by a small number of
features in the complex history of inputs. Following the ideas
which emerged in the analysis of motion sensitive neurons in the
fly \cite{bialek88,bill&robinprep},  \hfill\break we have identified
``features'' with ``dimensions'' and searched for low dimensional
descriptions of the input history which preserve the mutual
information between inputs and outputs (spike times). We have
considered only the generation of isolated spikes, leaving aside
the question of how spikes interact with one another as considered
by Kistler et al.. For these isolated spikes we began by searching
for projections onto a low dimensional linear subspace of the
originally $\sim 200$ dimensional stimulus space, and we found
that a substantial fraction of the mutual information could be
preserved in a model with just two dimensions.  Searching for the
information that is missing from this model, we found that rather
than adding more (Euclidean) dimensions we could capture $\sim
90\%$ of the information at high time resolution by keeping a two
dimensional description but allowing these dimensions to curve, so
that the neuron is sensitive to stimulus features which lie on a
two dimensional manifold.

The geometrical picture of neurons as being sensitive to features
that are defined by a low dimensional manifold in stimulus space
is attractive and, as noted in the Introduction, corresponds to a
widely shared intuition about the nature of neuronal feature
selectivity. While curved manifolds often appear as the targets
for learning in complex neural computations such as invariant
object recognition, the idea that such manifolds appear already in
the description of single neuron computation we believe to be
novel.

While we have exploited the fact that long simulations of the
Hodgkin--Huxley model are quite tractable to generate large
amounts of ``data'' for our analysis, it is important that, in the
end, our construction of a curved manifold as the relevant
stimulus subspace involves a series of computations which are just
simple generalizations of the conventional reverse correlation or
spike triggered average. This suggests that our approach can be
applied to real neurons without requiring qualitatively larger
data sets than might have been needed for a careful reverse
correlation analysis.  In the same spirit, recent work has shown
how covariance matrix analysis of the fly's motion sensitive
neurons can reveal nonlinear computations in a four dimensional
subspace using data sets of fewer than $10^4$ spikes
\cite{bill&robinprep}.  Low dimensional (linear) subspaces can
be found even in the response of model neurons to naturalistic
inputs if one searches directly for dimensions which capture the
largest fraction of the mutual information between inputs and
spikes \cite{tatyana}, and again the errors involved in
identifying the relevant dimensions are comparable to the errors
in reverse correlation \cite{tatyana2}. All of these results point
to the practical feasibility of describing real neurons in terms
of nonlinear computation on low dimensional relevant subspaces in
a high dimensional stimulus space.

Our reduced model of the Hodgkin--Huxley neuron both illustrates a
novel approach to dimensional reduction and gives new insight into
the computation performed by the neuron.  The reduced model is
essentially that of an edge detector for current trajectories, but
is sensitive to a further stimulus parameter, producing a curved
manifold. An interpretation of this curvature will be presented in
a forthcoming manuscript. This curved representation is able to
capture almost all information that isolated spikes convey about
the stimulus, or conversely, allow us to predict isolated spike
times with high temporal precision from the stimulus. The
emergence of a low dimensional curved manifold in a model as
simple as the Hodgkin--Huxley neuron suggests that such a
description may be also appropriate for biological neurons.

Our approach is limited in that we address only isolated spikes.
This restricted class of spikes nonetheless has biological
relevance; for example, in  vertebrate retinal ganglion cells
\cite{retinaISID} and in rat somatosensory cortex \cite{ras}, the
first spike of a burst has been shown to convey distinct (and the
majority of the) information.  However, a clear next step in this
program is to extend our formalism to take into account interspike
interaction. For neurons or models with explicit long timescales,
adaptation induces very long range history dependence which
complicates the issue of spike interactions considerably.  A full
understanding of the interaction between stimulus and spike
history will therefore, in general, involve understanding the
meanings of spike patterns \cite{bialek88,naama-info-method} and the
influence of the larger statistical context \cite{anature}. Our
results point to the need for a more parsimonious description of
self--excitation, even for the simple case of dependence only on
the last spike time.

We would like to close by reminding the reader of the more
ambitious goal of building bridges between the burgeoning
molecular level description of neurons and the functional or
computational level. Armed with a description of spike generation
as a nonlinear operation on a low dimensional, curved manifold in
the space of inputs, it is natural to ask how the details of this
computational picture are related to molecular mechanisms. Are
neurons with more different types of ion channels sensitive to
more stimulus dimensions, or do they implement more complex
nonlinearities in a low dimensional space? Are adaptation and
modulation mechanisms that change the nonlinearity separable from
those which change the dimensions to which the cell is sensitive?
Finally, while we have shown how a low dimensional description can
be constructed numerically from observations of the input/output
properties of the neuron, one would like to understand
analytically why such a description emerges and whether it emerges
universally from the combinations of channel dynamics selected by
real neurons.

\section*{Acknowledgments}

We thank N. Brenner for discussions at the start of this work, and
M. Berry for comments on the manuscript.


\bibliographystyle{apalike}
\bibliography{hhCurrentBill}

\begin{thebibliography}{}

\bibitem[Abbott and Kepler, 1990]{abbott}
Abbott, L.~F. and Kepler, T. (1990).
\newblock Model neurons: from hodgkin-huxley to hopfield.
\newblock In {\em Statistical Mechanics of Neural Networks}, pages 5--18,
  Berlin. Springer--Verlag.

\bibitem[Ag{\"u}era~y Arcas, 1998]{baathesis}
Ag{\"u}era~y Arcas, B. (1998).
\newblock Reducing the neuron: a computational approach.
\newblock Master's thesis, Princeton University.

\bibitem[Ag{\"u}era~y Arcas et~al., 2001]{nips2000}
Ag{\"u}era~y Arcas, B., Bialek, W., and Fairhall, A.~L. (2001).
\newblock What can a single neuron compute?
\newblock In Leen, T., Dietterich, T., and Tresp, V., editors, {\em Advances in
  Neural Information Processing Systems 13}, pages 75--81. MIT Press.

\bibitem[Ag{\"u}era~y Arcas and Fairhall, 2002]{i&f}
Ag{\"u}era~y Arcas, B. and Fairhall, A. (2002).
\newblock What causes a neuron to spike?
\newblock {\em submitted}.

\bibitem[Barlow, 1953]{barlow53}
Barlow, H.~B. (1953).
\newblock Summation and inhibition in the frog's retina.
\newblock {\em J. Physiol.}, 119:69--88.

\bibitem[Barlow et~al., 1964]{barlow2}
Barlow, H.~B., Hill, R.~M., and Levick, W.~R. (1964).
\newblock Retinal ganglion cells responding selectively to direction and speed
  of image motion in the rabbit.
\newblock {\em J. Physiol.}, 173:377--407.

\bibitem[Berry~II and Meister, 1999]{retinaISID}
Berry~II, M.~J. and Meister, M. (1999).
\newblock The neural code of the retina.
\newblock {\em Neuron}, 22:435--450.

\bibitem[Berry~II et~al., 1997]{berryWarlandMeister}
Berry~II, M.~J., Warland, D., and Meister, M. (1997).
\newblock The structure and precision of retinal spike trains.
\newblock {\em Proc. Natl. Acad. Sci. U.S.A.}, 94:5411--5416.

\bibitem[Bialek and de~Ruyter~van Steveninck, 2002]{bill&robinprep}
Bialek, W. and de~Ruyter~van Steveninck, R.~R. (2002).
\newblock Features and dimensions: motion estimation in fly vision.
\newblock {\em in preparation}.

\bibitem[Boser et~al., 1992]{svm1}
Boser, B.~E., Guyon, I.~M., and Vapnik, V.~N. (1992).
\newblock A training algorithm for optimal margin classifers.
\newblock In Haussler, D., editor, {\em 5th Annual ACM Workshop on COLT}, pages
  144--152, Pittsburgh, PA. ACM Press.

\bibitem[Bray, 1995]{bray}
Bray, D. (1995).
\newblock Protein molecules as computational elements in living cells.
\newblock {\em Nature}, 376:307--312.

\bibitem[Brenner et~al., 1998]{naamaISID}
Brenner, N., Agam, O., Bialek, W., and de~Ruy\-ter~van Steven\-inck, R.~R.
  (1998).
\newblock Universal statistical behavior of neural spike trains.
\newblock {\em Phys. Rev. Lett.}, 81:4000--4003.
\newblock See also Statistical properties of spike trains: universal and
  stimulus--dependent aspects, {\em Phys. Rev. E} 66, 031907 (2002);
  http://xxx.lanl.gov/abs/physics/9801026 and
  http://xxx.lanl.gov/abs/physics/9902061.

\bibitem[Brenner et~al., 2000a]{naama}
Brenner, N., Bialek, W., and de~Ruyter~van Steven\-inck, R.~R. (2000a).
\newblock Adaptive rescaling maximizes information transmission.
\newblock {\em Neuron}, 26:695--702.

\bibitem[Brenner et~al., 2000b]{naama-info-method}
Brenner, N., Strong, S., Koberle, R., Bialek, W., and de~Ruyter~van Steveninck,
  R.~R. (2000b).
\newblock Synergy in a neural code.
\newblock {\em Neural Comp.}, 12:1531--1552.
\newblock See also http://xxx.lanl.gov/abs/physics/9902067.

\bibitem[Cottrell et~al., 1988]{nonlinearPCA2}
Cottrell, G.~W., Munro, P., and Zipser, D. (1988).
\newblock Image compression by back propagation: a demonstration of extensional
  programming.
\newblock In Sharkey, N., editor, {\em Models of cognition: a review of
  cognitive science}, volume~2, pages 208--240, Norwood, NJ. Abbex.

\bibitem[Cover and Thomas, 1991]{cover&thomas}
Cover, T.~M. and Thomas, J.~A. (1991).
\newblock {\em Elements of Information Theory}.
\newblock John Wiley \& Sons, Inc., New York.

\bibitem[de~Boer and Kuyper, 1968]{deBoer}
de~Boer, E. and Kuyper, P. (1968).
\newblock Triggered correlation.
\newblock {\em IEEE Trans. Biomed. Eng.}, 15:169--179.

\bibitem[de~Ruy\-ter~van Steven\-inck and Bialek, 1988]{bialek88}
de~Ruy\-ter~van Steven\-inck, R.~R. and Bialek, W. (1988).
\newblock Real-time performance of a movement sensitive in the blowfly visual
  system: information transfer in short spike sequences.
\newblock {\em Proc. Roy. Soc. Lond. B}, 234:379--414.

\bibitem[de~Ruyter~van Steven\-inck et~al., 1997]{billscience}
de~Ruyter~van Steven\-inck, R., Lewen, G.~D., Strong, S.~P., Koberle, R., and
  Bialek, W. (1997).
\newblock Reproducibility and variability in neural spike trains.
\newblock {\em Science}, 275:1805--1808.

\bibitem[Dear et~al., 1993]{dear&simmons}
Dear, S.~P., Simmons, J.~A., and Fritz, J. (1993).
\newblock A possible neuronal basis for representation of acoustic scenes in
  auditory cortex of the big brown bat.
\newblock {\em Nature}, 364:620--623.

\bibitem[Fairhall et~al., 2001]{anature}
Fairhall, A., Lewen, G., Bialek, W., and de~Ruyter~van Steveninck, R.~R.
  (2001).
\newblock Efficiency and ambiguity in an adaptive neural code.
\newblock {\em Nature}, 412:787--792.

\bibitem[Fitzhugh, 1961]{fitzhugh}
Fitzhugh, R. (1961).
\newblock Impulse and physiological states in models of nerve membrane.
\newblock {\em Biophysics J.}, 1:445--466.

\bibitem[Guyon et~al., 1993]{svm2}
Guyon, I.~M., Boser, B.~E., and Vapnik, V.~N. (1993).
\newblock Automatic capacity tuning of very large vc-dimension classifiers.
\newblock In Hanson, S.~J., Cowan, J.~D., and Giles, C., editors, {\em Advances
  in Neural Information Processing Systems 5}, pages 147--155, San Mateo, CA.
  Morgan Kaufmann.

\bibitem[Hartline, 1940]{hartline}
Hartline, H.~K. (1940).
\newblock The receptive fields of optic nerve fibres.
\newblock {\em Amer. J. Physiol.}, 130:690--699.

\bibitem[Hille, 1992]{hille}
Hille, B. (1992).
\newblock {\em Ionic channels of excitable membranes.}
\newblock Sinaur, Sunderland, Massachusetts.

\bibitem[Hodgkin and Huxley, 1952]{hh52}
Hodgkin, A.~L. and Huxley, A.~F. (1952).
\newblock A quantitative description of membrane current and its application to
  conduction and excitation in nerve.
\newblock {\em J. Physiol.}, 463:391--407.

\bibitem[Hubel and Wiesel, 1962]{hubel&wiesel62}
Hubel, D.~H. and Wiesel, T.~N. (1962).
\newblock Receptive fields, binocular interaction and functional architecture
  in the cat's visual cortex.
\newblock {\em J. Physiol. (Lond.)}, 160:106--154.

\bibitem[Iverson and Zucker, 1995]{iverson&zucker}
Iverson, L. and Zucker, S.~W. (1995).
\newblock Logical/linear operators for image curves.
\newblock {\em IEEE Trans. Pattern Analysis and Machine Intelligence},
  17:982--996.

\bibitem[Keat et~al., 2001]{markusmodel}
Keat, J., Reinagel, P., Reid, R.~C., and Meister, M. (2001).
\newblock Predicting every spike: a model for the responses of visual neurons.
\newblock {\em Neuron}, 30(3):803--817.

\bibitem[Kepler et~al., 1992]{kepler}
Kepler, T., Abbott, L.~F., and Marder, E. (1992).
\newblock Reduction of conductance-based neuron models.
\newblock {\em Biological Cybernetics}, 66:381--387.

\bibitem[Kistler et~al., 1997]{gerstner}
Kistler, W., Gerstner, W., and van Hemmen, J.~L. (1997).
\newblock Reduction of the {H}odgkin-{H}uxley equations to a single-variable
  threshold model.
\newblock {\em Neural Computation}, 9:1015--1045.

\bibitem[Koch, 1999]{koch}
Koch, C. (1999).
\newblock {\em Biophysics of computation: information processing in single
  neurons}.
\newblock Oxford University Press, New York.

\bibitem[Kuffler, 1953]{kuffler}
Kuffler, S.~W. (1953).
\newblock Discharge patterns and functional organization of mammalian retina.
\newblock {\em J. Neurophysiol.}, 16:37--68.

\bibitem[Lewen et~al., 2001]{flyNaturalStim}
Lewen, G.~D., Bialek, W., and de~Ruyter~van Steveninck, R.~R. (2001).
\newblock Neural coding of naturalistic motion stimuli.
\newblock {\em Network}, 12:317--329.

\bibitem[Mainen and Sejnowski, 1995]{mainensejnowski}
Mainen, Z.~F. and Sejnowski, T.~J. (1995).
\newblock Reliability of spike timing in neocortical neurons.
\newblock {\em Science}, 268:1503--1506.

\bibitem[Nagumo et~al., 1962]{nagumo}
Nagumo, J., Arimoto, S., and Yoshikawa, Z. (1962).
\newblock An active pulse transmission line simulating nerve axon.
\newblock {\em Proc. IRE}, 50:2061--2071.

\bibitem[Oja and Karhunen, 1995]{nonlinearPCA1}
Oja, E. and Karhunen, J. (1995).
\newblock Signal separation by nonlinear hebbian learning.
\newblock In Palaniswami, M., Attikiouzel, Y., II, R.~M., Fogel, D., and
  Fukuda, T., editors, {\em Computational intelligence-- a dynamic system
  perspective}, pages 83--97, New York. IEEE Press.

\bibitem[Panzeri et~al., 2001]{ras}
Panzeri, S., Petersen, R., Schultz, S., Lebedev, M., and Diamond, M. (2001).
\newblock The role of spike timing in the coding of stimulus location in rat
  somatosensory cortex.
\newblock {\em Neuron}, 29:769--77.

\bibitem[Reinagel and Reid, 2000]{reinagel&reid}
Reinagel, P. and Reid, R.~C. (2000).
\newblock Temporal coding of visual information in the thalamus.
\newblock {\em J. Neuroscience}, 20(14):5392--5400.

\bibitem[Rieke et~al., 1997]{spikes}
Rieke, F., Warland, D., Bialek, W., and de~Ruyter~van Steveninck, R.~R. (1997).
\newblock {\em Spikes: exploring the neural code}.
\newblock The MIT Press, New York.

\bibitem[Rosenblatt, 1958]{rosen2}
Rosenblatt, F. (1958).
\newblock The perceptron: A probabilistic model for information storage and
  organization in the brain.
\newblock {\em Psychological Review}, 65:386--408.

\bibitem[Rosenblatt, 1962]{rosenblatt}
Rosenblatt, F. (1962).
\newblock {\em Principles of Neurodynamics}.
\newblock Spartan Books, New York.

\bibitem[Roweis and Saul, 2000]{roweis}
Roweis, S. and Saul, L. (2000).
\newblock Nonlinear dimensionality reduction by locally linear embedding.
\newblock {\em Science}, 290:2323--2326.

\bibitem[Schneidman et~al., 1998]{elad}
Schneidman, E., Freedman, B., and Segev, I. (1998).
\newblock Ion channel stochasticity may be critical in determing the
  reliability and precision of spike timing.
\newblock {\em Neural Comp.}, 10:1679--1703.

\bibitem[Shannon, 1948]{shannon}
Shannon, C.~E. (1948).
\newblock A mathematical theory of communication.
\newblock {\em Bell Sys. Tech. Journal}, 27:379--423, 623--656.

\bibitem[Sharpee et~al., 2002a]{tatyana}
Sharpee, T., Rust, N.~C., and Bialek, W. (2002a).
\newblock Maximally informative dimensions: analysing neural responses to
  natural signals.
\newblock {\em NIPS 2002 (submitted)}.

\bibitem[Sharpee et~al., 2002b]{tatyana2}
Sharpee, T., Rust, N.~C., and Bialek, W. (2002b).
\newblock Maximally informative dimensions: analysing neural responses to
  natural signals.
\newblock {\em in preparation}.

\bibitem[Stanley et~al., 1999]{dan}
Stanley, G.~B., Lei, F.~F., and Dan, Y. (1999).
\newblock Reconstruction of natural scenes from ensemble responses in the
  lateral geniculate nucleus.
\newblock {\em J. Neurosci.}, 19(18):8036--8042.

\bibitem[Theunissen et~al., 2000]{theunissen}
Theunissen, F., Sen, K., and Doupe, A. (2000).
\newblock Spectral-temporal receptive fields of nonlinear auditory neurons
  obtained using natural sounds.
\newblock {\em J. Neurosci.}, 20:2315--2331.

\bibitem[Tishby et~al., 1999]{bottleneck}
Tishby, N., Pereira, F., and Bialek, W. (1999).
\newblock The information bottleneck method.
\newblock In Hajek, B. and Sreenivas, R.~S., editors, {\em Proceedings of the
  37th Annual Allerton Conference on Communication, Control and Computing},
  pages 368--377. University of Illinois.
\newblock See also http://xxx.lanl.gov/abs/physics/0004057.

\end{thebibliography}

\end{document}